\documentclass[11pt]{article}
\usepackage{amssymb,amsmath,amsfonts}
\usepackage{graphicx}
\usepackage{graphics}
\usepackage{eepic,epsfig}


\begin{document}

\title{The Casimir densities for a sphere in the Milne universe}
\author{A. A. Saharian, T. A. Petrosyan \\
\\
\textit{Department of Physics, Yerevan State University,}\\
\textit{1 Alex Manoogian Street, 0025 Yerevan, Armenia}\vspace{0.3cm}\\
\textit{Institute of Applied Problems in Physics NAS RA,}\\
\textit{25 Nersessian Street, 0014 Yerevan, Armenia}}
\maketitle

\begin{abstract}
The influence of a spherical boundary on the vacuum fluctuations of a
massive scalar field is investigated in background of $(D+1)$-dimensional
Milne universe, assuming that the field obeys Robin boundary condition on
the sphere. The normalized mode functions are derived for the regions inside
and outside the sphere and different vacuum states are discussed. For the
conformal vacuum, the Hadamard function is decomposed into boundary-free and
sphere-induced contributions and an integral representation is obtained for
the latter in both the interior and exterior regions. As important local
characteristics of the vacuum state the vacuum expectation values (VEVs) of
the field squared and of the energy-momentum tensor are investigated. It is
shown that the vacuum energy-momentum tensor has an off-diagonal component
that corresponds to the energy flux along the radial direction. Depending on
the coefficient in Robin boundary condition the sphere-induced contribution
to the vacuum energy and the energy flux can be either positive or negative.
At late stages of the expansion and for a massive field the decay of the
sphere-induced VEVs, as functions of time, is damping oscillatory. The
geometry under consideration is conformally related to that for a static
spacetime with negative constant curvature space and the sphere-induced
contributions in the corresponding VEVs are compared.
\end{abstract}

\section{Introduction}

In constructing quantum field theories in background geometries different
from the Minkowski spacetime, among the most important points is the
selection of a physically meaningful vacuum state. The vacuum state depends
on the choice of complete set of mode functions used in the second
quantization procedure \cite{Birr82B}-\cite{Park09}. Those functions, and consequently the
properties of the vacuum state, are sensitive to both the local and global
characteristics of background spacetime. Already for the Minkowski bulk,
depending on spacetime coordinates employed corresponding to different
observers, different vacuum states are realized. An example is the
Fulling--Rindler vacuum that presents the vacuum state for a uniformly
accelerated observer different from the standard Minkowskian vacuum. The
corresponding coordinates (Rindler coordinates) cover only a part of the
Minkowski spacetime (R~and L Rindler wedges). Other patches of the Minkowski
spacetime inside the future and past light cones correspond to the so-called
Milne universe. Quantum field theory in the Milne patches of the Minkowski
spacetime has been discussed in \cite{Somm74}-\cite{Higu17} (see also \cite%
{Birr82B}-\cite{Park09}). The Milne universe is described by the
Friedmann--Robertson--Walker type line element with negative curvature spatial
sections and with the scale factor being a linear function of the time
coordinate. It is well adapted for considerations of different types of the
vacuum states in backgrounds with time-dependent metric tensors. Quantum
fields in background of closely related geometries with linear scale factors
and planar spatial sections have been considered in \cite{Full74}-\cite{Saha18}.

In the present paper we are interested in boundary-induced quantum effects
in the Milne universe. The boundary conditions imposed on quantum fields
modify the vacuum fluctuations and as a consequence of that the vacuum
expectation values (VEVs) of physical observables are shifted compared to
those for the geometry where the boundaries are absent. This is the
well-known Casimir effect (for reviews see \cite{Eliz94}-\cite{Dalv11}). Explicit
expressions for the vacuum characteristics can be obtained for highly
symmetric bulk and boundary geometries. In particular, spherical boundaries
have attracted a great deal of attention. The early investigations of the
electromagnetic Casimir effect for a reflecting spherical boundary were
motivated by the Casimir semiclassical model of an electron \cite{Casi53}
where the Casimir pressure compensates outwardly-directed coulomb forces.
However, the further studies of the Casimir effect for perfectly conducting
spherical shell \cite{Boye68}-\cite{Milt78} have shown that the vacuum forces are
outwardly-directed and, hence, cannot play the role of Poincar\'{e} stresses
(for further investigations of the Casimir energy and forces in geometries
with spherical boundaries see references in \cite{Eliz94}-\cite{Dalv11},\cite{Teo10,Leon11,Milt12}%
).

More detailed information on the properties of the vacuum state is provided
by local characteristics such as the VEVs of the field squared and of the
energy-momentum tensor. The latter VEV also determines the back reaction of
quantum effects on the background geometry and plays an important role in
modeling of self-consistent dynamics on the base of semiclassical Einstein
equations. The VEVs of the electric and magnetic field squared and of the
energy density for electromagnetic and chromomagnetic vacuum fields within a
spherical cavity with reflecting boundary have been studied in~\cite{Olau81,Olau81b}%
. The VEVs for the remaining components of the energy-momentum tensor for
the electromagnetic field inside and outside a spherical shell and in the
region between two concentric spherical shells is investigated in \cite%
{Brevik1}-\cite{Grig2} (these results are summarized in~\cite{Sahrev}). For a
massive scalar field with the Robin boundary condition the VEVs of the
energy-momentum tensor inside and outside a spherical shell and in the
region between two concentric spherical boundaries in $(D+1)$-dimensional
Minkowski spacetime have been investigated in \cite{Saha01}. The
corresponding VEVs in background of curved global monopole were discussed in
\cite{Saha04}-\cite{Saha04bd} for scalar and spinor fields, respectively.
The~Casimir densities in the geometry of a global monopole with a general
spherically symmetric static core of finite radius are investigated in \cite%
{Beze06,Beze07}. The bulk and surface Casimir densities for spherical branes in
Rindler-like spacetime $Ri\times S^{D-1}$, with a two-dimensional Rindler
spacetime $Ri$, have been studied in~\cite{RindSph,RindSph2,RindSph3}. The latter geometry
approximates the gravitational field near the horizon of $(D+1)$-dimensional
black holes. The VEV\ of the energy-momentum tensor for spherical boundaries
on the background of dS spacetime is investigated in \cite{Seta01,Seta01b} for a
conformally coupled massless scalar field and in \cite{Milt12} for a massive
field with general curvature coupling parameter and with the Robin boundary
condition. The scalar vacuum polarization and the expectation value of the
energy-momentum tensor for spherical boundaries in the background of a
constant negative curvature space are discussed in \cite{Saha08,Saha08b,Bell14}. In
\cite{Bell14b} the two-point function and the VEVs are investigated for a
scalar field in a spherically symmetric static background geometry with
two distinct metric tensors inside and outside a spherical boundary. In
this setup the exterior and interior geometries can correspond to different
vacuum states of the same theory.

Here we investigate the local characteristics of the scalar vacuum inside
and outside a spherical boundary in the Milne universe. The corresponding
line element is conformally related to that for a static spacetime with
constant negative curvature space with time-dependent conformal factor.
Though the geometry of the Milne universe is flat, related to the
time-dependence of the metric tensor, the~Casimir problem for a spherical
boundary in its background is more complicated than that discussed in \cite%
{Saha08,Bell14} for a curved background.

The paper is organized as follows. In the next section we describe the bulk
and boundary geometries and the structure of the modes for a scalar field
with Robin boundary condition on a sphere. The mode functions are specified
for the adiabatic and conformal vacua. In Section \ref{sec:int} the Hadamard
function, the boundary-induced contributions in the VEVs of the field
squared and of the energy-momentum tensor are investigated inside the
spherical shell for the conformal vacuum. The behavior of the VEVs in
asymptotic regions of the parameters are discussed and the results of the
numerical analysis are presented. The similar investigation for the region
outside the sphere is presented in Section \ref{sec:Ext}. The main results
of the paper are summarized in Section \ref{sec:Conc}. In Appendix \ref%
{sec:appSF} we present a summation formula for series over the scalar
eigenmodes inside the spherical shell that is used to derive an integral
representation for the Hadamard function.

\section{Geometry and the Scalar Field Modes}

\label{sec:Geom}

The $(D+1)$-dimensional background spacetime we are going to consider is
described by the line~element%
\begin{equation}
ds^{2}=dt^{2}-t^{2}(dr^{2}+\sinh ^{2}rd\Omega _{D-1}^{2}),  \label{ds2}
\end{equation}%
where $0\leq t<\infty $, $0\leq r<\infty $, and $d\Omega _{D-1}^{2}$ is the
line element on a sphere $S^{D-1}$ with unit radius. The~spatial part of the
line element is written in terms of the hyperspherical coordinates $%
(r,\vartheta ,\phi )$ with the set of angular coordinates $\vartheta
=(\theta _{1},\theta _{2},\ldots \theta _{n})$, where $n=D-2$. Note that the
radial coordinate $r$ is dimensionless. The line element (\ref{ds2})
corresponds to the so-called Milne universe. The background spacetime is
flat, however the spatial geometry corresponds to a negatively curved space.

\begin{figure}[tbph]
\begin{center}
\epsfig{figure=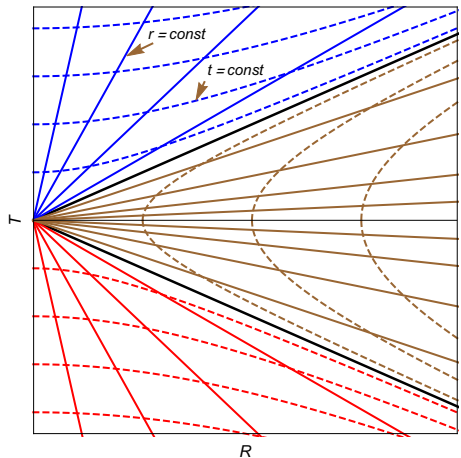,width=6cm,height=6cm}
\end{center}
\caption{Milne and Rindler patches of the Minkowski spacetime in the $(T,R)$
half-plane.}
\label{fig1}
\end{figure}

The flatness of the geometry described by (\ref{ds2}) is explicitly seen
passing to new coordinates $(T,R,\vartheta ,\phi )$ with
\begin{equation}
T=t\cosh r,\;R=t\sinh r.  \label{TR}
\end{equation}%
In terms of these coordinates the line element takes the standard
Minkowskian form in spherical spatial coordinates:
\begin{equation}
ds^{2}=dT^{2}-dR^{2}-R^{2}d\Omega _{D-1}^{2}.  \label{ds2M}
\end{equation}%
From (\ref{TR}) we see that the coordinates $(t,r,\vartheta ,\phi )$ cover
the patch of the Minkowski spacetime inside the future light cone. In Figure %
\ref{fig1} we have plotted the coordinate lines for $(t,r)$ in the
Minkowskian half-plane $(T,R)$. In the past light cone one has $T=t\cosh r$,
$R=-t\sinh r$, $-\infty <t\leq 0$. The~remaining region $R>|T|$ corresponds
to the Rindler patch. From the cosmological point of view, the importance of
the metric given by (\ref{ds2}) is related to the fact that it is the late
time attractor in a large class of Friedmann--Robertson--Walker open
cosmological models.

Consider a massive scalar field $\varphi (x)$ with a curvature coupling
parameter $\xi $. The corresponding equation of motion reads
\begin{equation}
(\nabla _{\mu }\nabla ^{\mu }+m^{2}+\xi \mathcal{R})\varphi (x)=0,
\label{Feq}
\end{equation}%
where $\nabla _{\mu }$ is the covariant derivative operator and $\mathcal{R}$
is the Ricci scalar for the background geometry. Though in the problem under
consideration $\mathcal{R}=0$, the energy-momentum tensor depends on the
curvature coupling parameter. We assume the presence of a spherical boundary
with radius $r_{0}$ on which the scalar field obeys Robin boundary condition%
\begin{equation}
\left( A-\delta _{(j)}B\partial _{r}\right) \varphi (x)=0,\;r=r_{0},
\label{Robin}
\end{equation}%
where $A$ and $B$ are dimensionless constants, $j=i,e$, with $\delta
_{(i)}=1 $ for the interior region and $\delta _{(e)}=-1$ for the exterior
region. Special cases $B=0$ and $A=0$ correspond to Dirichlet and Neumann
boundary conditions. Let us denote by $R_{0}$ the radius of the sphere in
terms of the Minkowskian radial coordinate $R$. From (\ref{TR}) we get $%
R_{0}=T\tanh r_{0}$. Hence, the boundary under consideration corresponds to
a uniformly expanding sphere in the Minkowski spacetime with the velocity $%
\tanh r_{0}$. We are interested in the influence of the sphere on the local
properties of the vacuum state. Those properties are encoded in two-point
functions and, as the first step, we shall evaluate these functions. In
order to do that we need a complete set of mode functions for the field
obeying the boundary condition (\ref{Robin}).

In the hyperspherical coordinates the field equation is written as%
\begin{equation}
\left[ \frac{1}{t^{D}}\partial _{t}\left( t^{D}\partial _{t}\right) -\frac{%
\partial _{1}\left( \sinh ^{D-1}r\partial _{r}\right) }{t^{2}\sinh ^{D-1}r}-%
\frac{\Delta _{\vartheta ,\phi }}{t^{2}\sinh ^{2}r}+m^{2}\right] \varphi
(x)=0,  \label{Feq2}
\end{equation}%
where $\Delta _{\vartheta ,\phi }$ is the angular part of the Laplacian
operator. In accordance with the problem symmetry, we present the solution
in the form%
\begin{equation}
\varphi \left( t,r,\vartheta ,\phi \right) =f(t)g(r)Y(m_{p};\vartheta ,\phi
),  \label{phisol}
\end{equation}%
where $Y(m_{p};\vartheta ,\phi )$ are the hyperspherical harmonics of degree
$l$, where $m_{p}=(m_{0}\equiv l,m_{1},\ldots ,m_{n})$, $l=0,1,2,\ldots $,
and $m_{1},m_{2},\ldots ,m_{n}$ are integers such that $-m_{n-1}\leqslant
m_{n}\leqslant m_{n-1}$ and
\begin{equation}
0\leqslant m_{n-1}\leqslant m_{n-2}\leqslant \cdots \leqslant m_{1}\leqslant
l.  \label{mn-1}
\end{equation}%
The hyperspherical harmonics obey the equation
\begin{equation}
\Delta _{\vartheta ,\phi }Y(m_{p};\vartheta ,\phi )=-l(l+n)Y(m_{p};\vartheta
,\phi ),  \label{EqY}
\end{equation}%
and are normalized in accordance with%
\begin{equation}
\int d\Omega \,\left\vert Y(m_{p};\vartheta ,\phi )\right\vert ^{2}=N(m_{p}).
\label{Yint}
\end{equation}%
The explicit expression for $N(m_{p})$ is not required in the following
discussion. One has the addition~theorem%
\begin{equation}
\sum_{m_{p}}\frac{Y(m_{p};\vartheta ,\phi )}{N(m_{p})}Y^{\ast
}(m_{p};\vartheta ^{\prime },\phi ^{\prime })=\frac{2l+n}{nS_{D}}%
C_{l}^{n/2}(\cos \theta ),  \label{Addth}
\end{equation}%
where the star means the complex conjugate, $\theta $ is the
angle between the directions determined by $(\vartheta ,\phi )$ and $%
(\vartheta ^{\prime },\phi ^{\prime })$, $S_{D}=2\pi ^{D/2}/\Gamma
(D/2)$ is the surface area of the sphere in $D$-dimensional space with unit radius, and $%
C_{l}^{n/2}(\cos \theta )$ is the Gegenbauer polynomial.

By taking into account (\ref{EqY}), the Equation (\ref{Feq2}) is rewritten as%
\begin{equation}
\frac{\partial _{t}\left( t^{D}\partial _{t}f(t)\right) }{f(t)t^{D-2}}%
+t^{2}m^{2}-\frac{\partial _{1}\left( \sinh ^{D-1}r\partial _{r}g(r)\right)
}{g(r)\sinh ^{D-1}r}+\frac{l(l+n)}{\sinh ^{2}r}=0.  \label{Feq3}
\end{equation}%
From here two separate equations are obtained:%
\begin{eqnarray}
\partial _{t}^{2}f(t)+\frac{D}{t}\partial _{t}f(t)+\left( m^{2}+\frac{\gamma
^{2}}{t^{2}}\right) f(t) &=&0,  \notag \\
\frac{\partial _{r}\left( \sinh ^{D-1}r\partial _{r}g(r)\right) }{\sinh
^{D-1}r}+\left[ \gamma ^{2}-\frac{l(l+n)}{\sinh ^{2}r}\right] g(r) &=&0,
\label{Eqg}
\end{eqnarray}%
where $\gamma $ is the separation constant. Introducing a new radial
function $h(r)=\sinh ^{D/2-1}(r)g(r)$, we~can see that the general solution
for $h(r)$ is a linear combination of the associated Legendre functions $P_{iz-1/2}^{-\mu }(\cosh r)$ and $%
Q_{iz-1/2}^{-\mu }(\cosh r)$ (here the associated Legendre
functions are defined in accordance with Ref. \cite{Abra72}) with the order and degree
determined by%
\begin{equation}
\mu =l+D/2-1,\;z^{2}=\gamma ^{2}-(D-1)^{2}/4.  \label{mu}
\end{equation}%
The relative coefficient in the linear combination depends on the spatial
region under consideration and will be determined below.

The general solution for the function $f(t)$ is presented in two equivalent
forms%
\begin{eqnarray}
f(t) &=&t^{(1-D)/2}\left[ d_{1}J_{-iz}(mt)+d_{2}J_{iz}(mt)\right]  \notag \\
&=&t^{(1-D)/2}\left[ c_{1}H_{iz}^{(1)}(mt)+c_{2}H_{iz}^{(2)}(mt)\right] ,
\label{fsol}
\end{eqnarray}%
with constants $d_{1,2}$ and $c_{1,2}$. The coefficients in two
representations are related by%
\begin{equation}
c_{1}=\frac{e^{-z\pi }d_{1}+d_{2}}{2},\;c_{2}=\frac{e^{z\pi }d_{1}+d_{2}}{2}.
\label{cd}
\end{equation}%
We will assume that the function $f(t)$ is normalised by the condition%
\begin{equation}
f^{\ast }(t)f^{\prime }(t)-f(t)f^{\ast \prime }(t)=-it^{-D}.  \label{fnorm}
\end{equation}%
This leads to the following relations between the coefficients $c_{1}$ and $%
c_{2}$:%
\begin{equation}
e^{\pi (z+z^{\ast })/2}|c_{2}|^{2}-e^{-\pi (z+z^{\ast })/2}|c_{1}|^{2}=\frac{%
\pi }{4}.  \label{Relc12}
\end{equation}%
For real $z$ the corresponding relation between the coefficients $d_{1}$ and
$d_{2}$ has the form%
\begin{equation}
|d_{1}|^{2}-|d_{2}|^{2}=\frac{\pi }{2\sinh (z\pi )},  \label{Reld12}
\end{equation}%
whereas for purely imaginary $z$ we get%
\begin{equation}
d_{1}d_{2}^{\ast }-d_{1}^{\ast }d_{2}=\frac{\pi }{2\sinh (z\pi )}.
\label{Reld12b}
\end{equation}

Hence, the mode functions for the scalar field are written as%
\begin{equation}
\varphi _{\sigma }\left( x\right) =\frac{X_{iz}(mt)}{t^{(D-1)/2}}\frac{%
Z_{iz-1/2}^{-\mu }(u)}{\sinh ^{D/2-1}r}Y(m_{p};\vartheta ,\phi ),
\label{phi}
\end{equation}%
where $\sigma $ stands for the set of quantum numbers specifying the
solutions, $x=\left( t,r,\vartheta ,\phi \right) $, and%
\begin{eqnarray}
X_{iz}(mt) &=&d_{1}J_{-iz}(mt)+d_{2}J_{iz}(mt),  \notag \\
Z_{iz-1/2}^{-\mu }(u) &=&b_{1}P_{iz-1/2}^{-\mu }(u)+b_{2}Q_{iz-1/2}^{-\mu
}(u),  \label{XZ}
\end{eqnarray}%
with $u=\cosh r$. By taking into account the condition (\ref{fnorm}) imposed
on the function $f(t)$, from the normalization condition
\begin{equation}
-i\int d^{D}x\sqrt{|g|}\varphi _{\sigma }(x)\overleftrightarrow{\partial }%
_{t}\varphi _{\sigma ^{\prime }}^{\ast }(x)=\delta _{\sigma \sigma ^{\prime
}},
\end{equation}%
we get the corresponding condition for the radial function:%
\begin{equation}
\int du\,Z_{iz-1/2}^{-\mu }(u)[Z_{iz^{\prime }-1/2}^{-\mu }(u)]^{\ast }=%
\frac{\delta _{zz^{\prime }}}{N(m_{p})},  \label{norm2}
\end{equation}%
where the integration goes inside or outside the sphere. After imposing the
boundary condition, the~normalized mode functions contain an arbitrary
constant that should be fixed by the choice of the vacuum state.

In order to discuss the set of vacuum states it is convenient to introduce
new coordinates $(\eta ,\bar{r})$ according to%
\begin{equation}
t=ae^{\eta /a},\;r=\bar{r}/a,  \label{Coord}
\end{equation}%
where $a$ is a constant with dimension of length. In terms of these
coordinates the line element is written in conformally static form%
\begin{equation}
ds^{2}=e^{2\eta /a}\left[ d\eta ^{2}-d\bar{r}^{2}-a^{2}\sinh ^{2}(\bar{r}%
/a)d\Omega _{D-1}^{2}\right] .  \label{ds2n}
\end{equation}%
This shows that the background geometry under consideration is conformally
related (with the conformal factor $(t/a)^{2}$) to the static spacetime with
a constant negative curvature space. In the limit $a\rightarrow \infty $ for
fixed $\eta $ and $\bar{r}$, from (\ref{ds2n}) the Minkowskian line element
in spherical coordinates is obtained. In this limit the quantum number $%
\gamma $ appears in equations (\ref{Eqg}), written in terms of the
coordinates $(\eta ,\bar{r})$, in the form of the ratio $\gamma /a$. The
Minkowskian mode functions are obtained for fixed $w=\gamma /a$. This~means
that in the limit $a\rightarrow \infty $ one has $\gamma =aw$, with fixed $w$%
, and $\gamma $ tends to infinity. The influence of a spherical boundary on
the properties of the vacuum state for static spacetime with a negative
constant curvature space (the corresponding line element is given by the
expression in the square brackets of (\ref{ds2n})) has been investigated in
\cite{Saha08,Bell14}.

Let us consider the function (\ref{fsol}) in the Minkowskian limit $%
a\rightarrow \infty $. Both the order and the argument of the cylindrical
functions are large and we use the leading terms of the corresponding
uniform asymptotic expansions. For the Hankel functions that gives%
\begin{equation*}
e^{z\pi /2}H_{iz}^{(2)}(mt)\approx \sqrt{\frac{2a}{\pi E}}e^{-i\gamma \xi
(m/w)+\pi i/4}e^{-iE\eta },
\end{equation*}%
and $e^{-z\pi /2}H_{iz}^{(1)}(mt)=[e^{z\pi /2}H_{iz}^{(2)}(mt)]^{\ast }$,
with $E=\sqrt{w^{2}+m^{2}}$ and
\begin{equation}
\xi (x)=\sqrt{1+x^{2}}+\ln \frac{x}{1+\sqrt{1+x^{2}}}.  \label{ksi}
\end{equation}%
By taking into account that to the leading order $t\approx a$, we see that
from the part in (\ref{fsol}) with the Hankel function $H_{iz}^{(2)}(mt)$
the Minkowskian positive energy mode functions are obtained. Hence, in order
to obtain the Minkowskian vacuum one should take in (\ref{fsol}) $c_{1}=0$.
This corresponds to the adiabatic vacuum, denoted here as $|0_{A}\rangle $
(see also the discussion in \cite{Birr82B,Full74}). For the corresponding
function $X_{iz}(mt)$ in (\ref{phi}) we have%
\begin{equation}
X_{iz}(mt)=\frac{\sqrt{\pi }}{2}e^{z\pi /2}H_{iz}^{(2)}(mt).  \label{Xadn}
\end{equation}

In accordance with (\ref{ds2n}) for a conformally coupled massless scalar
field the problem under consideration is conformally related to the
corresponding problem in static spacetime with a constant negative curvature
space. In the limit $m\rightarrow 0$ from (\ref{XZ}) one has%
\begin{equation}
X_{iz}(mt)=\frac{d_{1}e^{-iz\ln (ma/2)}}{\Gamma (1-iz)}e^{-iz\eta /a}+\frac{%
d_{2}e^{iz\ln (ma/2)}}{\Gamma (1+iz)}e^{iz\eta /a},  \label{Xc}
\end{equation}%
where $\Gamma (x)$ is the gamma function. For the case $d_{2}=0$ the mode
functions (\ref{phi}) are conformally related to the corresponding positive
energy mode functions in the static counterpart (with the energy $E=z/a$).
From the relation (\ref{Reld12b}) it follows that the quantum number $z$
should be real for the modes realizing the corresponding vacuum. The latter
is called as the conformal vacuum. The coefficient $d_{1}$ is found from (%
\ref{Reld12}) and the mode functions have the form (\ref{phi}) with
\begin{equation}
X_{iz}(mt)=\sqrt{\frac{\pi }{2\sinh (\pi z)}}J_{-iz}(mt).  \label{Xconf}
\end{equation}%
In what follows we will assume that the filed is prepared in the conformal
vacuum. The mode function are different in the exterior and interior region
of the sphere and we consider them separately.

\section{Region inside the Sphere}

\label{sec:int}

\subsection{Normalized Mode Functions}

Inside the sphere, $r<r_{0}$, for the modes (\ref{phi}) regular at the
sphere center, in (\ref{XZ}) one has $b_{2}=0$. The mode functions realizing
the conformal vacuum take the form
\begin{equation}
\varphi _{\sigma }\left( x\right) =C_{\sigma }^{(\mathrm{i})}\frac{X_{iz}(mt)%
}{t^{(D-1)/2}}\frac{P_{iz-1/2}^{-\mu }(u)}{\sinh ^{D/2-1}r}Y(m_{p};\vartheta
,\phi ),  \label{phii}
\end{equation}%
where $X_{iz}(mt)$ is given by (\ref{Xconf}). As it has been explained in
the previous section, in (\ref{phii}) $0\leq z<\infty $. The corresponding
eigenvalues are determined by the boundary condition (\ref{Robin}). They are
roots of the~equation
\begin{equation}
\bar{P}_{iz-1/2}^{-\mu }(u_{0})=0,\;u_{0}=\cosh r_{0}.  \label{Eigeq}
\end{equation}%
For a given function $f(x)$, the barred notation in the right-hand side is
defined in accordance with
\begin{equation}
\bar{f}(x)=A(x)f(x)+B(x)f^{\prime }(x),  \label{Barnot}
\end{equation}%
with the coefficients%
\begin{equation}
A(x)=A\sqrt{x^{2}-1}+(D/2-1)\delta _{(j)}Bx,\;B(x)=-\delta _{(j)}B(x^{2}-1).
\label{Au}
\end{equation}%
The corresponding positive solutions will be denoted by $z=z_{k}$, $%
k=1,2,\ldots $. The set of quantum numbers are specified by $\sigma
=(k,m_{p})$.

The Equation (\ref{Eigeq}) coincides with the eigenvalue equation inside a
spherical boundary in static spacetime with a negative constant curvature
space and the properties of the roots were discussed in \cite{Bell14}. It
has been shown that for $\beta \equiv A/B\geq -(D-1)/2$ in addition to real
roots the Equation~(\ref{Eigeq}) may have purely imaginary solutions. With
given $l$ and $u_{0}$ there are no purely imaginary zeros for sufficiently
small values of $\beta $. With increasing $\beta $, started from some
critical value $\beta =\beta _{l}^{(\mathrm{i})}(u_{0})$, purely~imaginary
zeros $z=\pm i\eta _{l}$, $\eta _{l}>0$, appear. The critical value $\beta
_{l}^{(\mathrm{i})}(u_{0})$ increases when $l$ increases and the purely
imaginary roots first appear for the lowest mode $l=0$. For a given $u_{0}$
there are no purely imaginary zeros if $\beta <\beta _{0}^{(\mathrm{i}%
)}(u_{0})$. For the critical values of the Robin coefficient one has $\beta
_{l}^{(\mathrm{i})}(u_{0})>-(D-1)/2$ and they are decreasing function of $%
u_{0}$. In Table \ref{tab1}, for the spatial dimension $D=3$, we present the
critical values of the Robin coefficient in the interior region for the mode
$l=0$ and for different values of the sphere radius. As it has been
discussed above, for the conformal vacuum the quantum number $z$ should be
real. Related to this, in what follows we will assume the values of $\beta $
for which all the roots of (\ref{Eigeq}) are real.

\begin{table}[tbph]
\caption{The critical values of the Robin coefficient for $l=0$ versus the
sphere radius.}
\label{tab1}\centering
\begin{tabular}{ccccccccc}
\hline
$r_{0}$ & 0.5 & 1 & 1.5 & 2 & 4 & 6 & 8 & 10 \\
$-\beta _{0}^{(\mathrm{i})}(u_{0})$ & 0.164 & 0.313 & 0.438 & 0.537 & 0.751
& 0.833 & 0.875 & 0.9 \\ \hline
\end{tabular}%
\end{table}

For $z^{\prime }=z$ the integral in the normalization condition (\ref{norm2}%
) is reduced to $\int_{1}^{u_{0}}du\,[P_{iz-1/2}^{-\mu }(u)]^{2}$. By~taking
into account that $z=z_{k}$ are the roots of (\ref{Eigeq}), the integral can
be presented in the form
\begin{equation}
\int_{1}^{u_{0}}du\,[P_{iz-1/2}^{-\mu }(u)]^{2}=\frac{1}{2zB}%
P_{iz-1/2}^{-\mu }(u_{0})\partial _{z}\bar{P}_{iz-1/2}^{-\mu }(u_{0}).
\label{P2int}
\end{equation}%
For the further consideration it is convenient to introduce the function%
\begin{eqnarray}
T_{\mu }(z,u_{0}) &=&\frac{\pi e^{-i\mu \pi }B|\Gamma (\mu +iz+1/2)|^{-2}}{%
P_{iz-1/2}^{-\mu }(u_{0})\partial _{z}\bar{P}_{iz-1/2}^{-\mu }(u_{0})}
\notag \\
&=&\frac{\bar{Q}_{iz-1/2}^{-\mu }(u_{0})}{\partial _{z}\bar{P}%
_{iz-1/2}^{-\mu }(u_{0})}\cos [\pi (\mu -iz)],  \label{T2}
\end{eqnarray}%
for $z=z_{k}$. In the second representation we have used the relation%
\begin{equation}
\bar{Q}_{iz-1/2}^{-\mu }(u_{0})=\frac{Be^{-i\mu \pi }\Gamma (iz-\mu +1/2)}{%
\Gamma (iz+\mu +1/2)P_{iz-1/2}^{-\mu }(u_{0})},  \label{QW}
\end{equation}%
valid for $z=z_{k}$. With the notation (\ref{T2}), the normalization
constant is presented as%
\begin{equation}
|C_{\sigma }^{(\mathrm{i})}|^{2}=e^{i\mu \pi }\frac{2z_{k}T_{\mu
}(z_{k},u_{0})}{\pi N(m_{p})}|\Gamma (\mu +iz_{k}+1/2)|^{2},  \label{b1n}
\end{equation}%
where $\mu $ is given by (\ref{mu}). Having specified the scalar modes, we
turn to the evaluation of the Hadamard function inside the sphere.

\subsection{Hadamard Function}

The Hadamard function is defined as the VEV $G(x,x^{\prime })=\langle
0|\varphi (x)\varphi (x^{\prime })+\varphi (x^{\prime })\varphi (x)|0\rangle
$. Expanding the field operators in terms of the mode functions $\varphi
_{\sigma }(x)$ and using the commutation relations, it is presented in the
form of the mode sum%
\begin{equation}
G(x,x^{\prime })=\sum_{k=1}^{\infty }\sum_{m_{p}}\left[ \varphi _{\sigma
}(x)\varphi _{\sigma }^{\ast }(x^{\prime })+\varphi _{\sigma }(x^{\prime
})\varphi _{\sigma }^{\ast }(x)\right] .  \label{WFsum}
\end{equation}%
With the mode functions from (\ref{phii}), for the Hadamard function one gets%
\begin{eqnarray}
G(x,x^{\prime }) &=&\frac{\left( tt^{\prime }\right) ^{(1-D)/2}}{nS_{D}}%
\sum_{l=0}^{\infty }\frac{\left( 2l+n\right) C_{l}^{n/2}(\cos \theta )}{%
(\sinh r\sinh r^{\prime })^{D/2-1}}e^{i\mu \pi }\sum_{k=1}^{\infty }zT_{\mu
}(z,u_{0})  \notag \\
&&\times |\Gamma (\mu +iz+1/2)|^{2}W(t,t^{\prime },z)P_{iz-1/2}^{-\mu
}(u)P_{iz-1/2}^{-\mu }(u^{\prime })|_{z=z_{k}},  \label{Wi}
\end{eqnarray}%
where $u^{\prime }=\cosh r^{\prime }$ and we have introduced the notation%
\begin{equation}
W(t,t^{\prime },z)=\frac{J_{-iz}(mt)J_{iz}(mt^{\prime
})+J_{iz}(mt)J_{-iz}(mt^{\prime })}{\sinh (z\pi )}.  \label{W}
\end{equation}%
In (\ref{Wi}), the summation over the roots of (\ref{Eigeq}) is present.
These roots are given implicitly and the representation (\ref{Wi}) is not
convenient for the investigation of the local characteristics of the vacuum~state.

A representation more adapted for the further analysis is obtained by using
the summation formula (\ref{SumForm}) from Appendix \ref{sec:appSF}. For the
series in (\ref{Wi}) the corresponding function is given by%
\begin{equation}
h(z)=z\Gamma (\mu +iz+1/2)\Gamma (\mu -iz+1/2)P_{iz-1/2}^{-\mu
}(u)P_{iz-1/2}^{-\mu }(u^{\prime })W(t,t^{\prime },z).  \label{hz}
\end{equation}%
This function has simple poles at $z=\pm ik$, $k=1,2,\ldots $. They
correspond to the poles $\pm ix_{k}$ discussed in Appendix \ref{sec:appSF}.
By taking into account the relation $P_{-x-1/2}^{-\mu }(u)=P_{x-1/2}^{-\mu
}(u)$, we see that the function (\ref{hz})\ is an even function. Hence, the
term in (\ref{SumForm}) containing the residues at the poles $\pm ix_{k}$
becomes zero. In the last integral of (\ref{SumForm}) we use the relation
\cite{Abra72}%
\begin{equation}
Q_{x-1/2}^{-\mu }(u)=e^{-2i\mu \pi }\frac{\Gamma (x-\mu +1/2)}{\Gamma (x+\mu
+1/2)}Q_{x-1/2}^{\mu }(u).  \label{RelQ}
\end{equation}

Applying the summation formula (\ref{SumForm}) with the function (\ref{hz}),
the Hadamard function is presented~as
\begin{equation}
G(x,x^{\prime })=G_{0}(x,x^{\prime })+G_{\mathrm{b}}(x,x^{\prime }),
\label{Wdec}
\end{equation}%
where the part%
\begin{eqnarray}
G_{0}(x,x^{\prime }) &=&\frac{\left( tt^{\prime }\right) ^{(1-D)/2}}{2nS_{D}}%
\sum_{l=0}^{\infty }\frac{\left( 2l+n\right) C_{l}^{n/2}(\cos \theta )}{%
(\sinh r\sinh r^{\prime })^{D/2-1}}\int_{0}^{\infty }dx\,x\sinh (\pi x)
\notag \\
&&\times |\Gamma (\mu +ix+1/2)|^{2}W(t,t^{\prime },x)P_{ix-1/2}^{-\mu
}(u)P_{ix-1/2}^{-\mu }(u^{\prime }),  \label{W0}
\end{eqnarray}%
is the corresponding function in the boundary-free geometry. The last term
in (\ref{Wdec}) comes from the second integral in (\ref{SumForm}) and is
expressed as%
\begin{eqnarray}
G_{\mathrm{b}}(x,x^{\prime }) &=&-\frac{\left( tt^{\prime }\right) ^{(1-D)/2}%
}{nS_{D}}\sum_{l=0}^{\infty }\frac{\left( 2l+n\right) C_{l}^{n/2}(\cos
\theta )}{(\sinh r\sinh r^{\prime })^{D/2-1}}e^{-i\mu \pi }  \notag \\
&&\times \int_{0}^{\infty }dx\,x\frac{\bar{Q}_{x-1/2}^{\mu }(u_{0})}{\bar{P}%
_{x-1/2}^{-\mu }(u_{0})}V(t,t^{\prime },x)P_{x-1/2}^{-\mu
}(u)P_{x-1/2}^{-\mu }(u^{\prime }),  \label{Wb}
\end{eqnarray}%
where%
\begin{equation}
V(t,t^{\prime },x)=\frac{J_{x}(mt)J_{-x}(mt^{\prime
})+J_{-x}(mt)J_{x}(mt^{\prime })}{\sin (\pi x)},  \label{Vtt}
\end{equation}%
and the integral is understood in the sense of the principal value.

For a massless field one has
\begin{equation}
\lim_{m\rightarrow 0}V(t,t^{\prime },x)=\frac{2}{\pi x}\cosh \left[ x\left(
\eta -\eta ^{\prime }\right) /a\right] ,  \label{m0}
\end{equation}%
with the conformal time $\eta $ from (\ref{Coord}). In this case the
sphere-induced contribution to the Hadamard function is transformed to%
\begin{equation}
G_{\mathrm{b}}(x,x^{\prime })=e^{(1-D)(\eta +\eta ^{\prime })/(2a)}G_{%
\mathrm{b}}^{(\mathrm{st})}(x,x^{\prime }),  \label{Gst}
\end{equation}%
where%
\begin{eqnarray}
G_{\mathrm{b}}^{(\mathrm{st})}(x,x^{\prime }) &=&-\frac{2}{a^{D-1}}%
\sum_{l=0}^{\infty }\frac{2l+n}{\pi nS_{D}}e^{-i\mu \pi }C_{l}^{n/2}(\cos
\theta )\int_{0}^{\infty }dx\,\frac{\bar{Q}_{x-1/2}^{\mu }(u_{0})}{\bar{P}%
_{x-1/2}^{-\mu }(u_{0})}  \notag \\
&&\times \frac{P_{x-1/2}^{-\mu }(u)P_{x-1/2}^{-\mu }(u^{\prime })}{(\sinh
r\sinh r^{\prime })^{D/2-1}}\cosh \left( \frac{\eta -\eta ^{\prime }}{a}%
x\right) ,  \label{Gstb}
\end{eqnarray}%
is the Hadamard function for a conformally coupled massless field in a
static spacetime with a negative constant curvature space (see \cite{Bell14}%
). The line element for the latter geometry is given by the expression in
the square brackets of (\ref{ds2n}) and $r=\bar{r}/a$:%
\begin{equation}
ds_{(\mathrm{st})}^{2}=d\eta ^{2}-a^{2}\left( dr^{2}+\sinh ^{2}rd\Omega
_{D-1}^{2}\right) .  \label{ds2st}
\end{equation}%
Note that for the corresponding Ricci scalar one has $\mathcal{R}%
=-D(D-1)/a^{2}$. From (\ref{Gst}) we see the conformal relation between the
problems in the Milne universe and in the static spacetime (\ref{ds2st}). Of~course, that is a consequence of our choice of the conformal vacuum.

\subsection{VEV of the Field Squared}

The VEV of the field squared is obtained from the Hadamard function in the
coincidence limit of the arguments. That limit is divergent and a
renormalization is required. For points away from the sphere the divergences
are contained in the boundary-free part only. The VEV is decomposed as%
\begin{equation}
\left\langle \varphi ^{2}\right\rangle =\left\langle \varphi
^{2}\right\rangle _{0}+\left\langle \varphi ^{2}\right\rangle _{\mathrm{b}},
\label{phi2dec}
\end{equation}%
where $\left\langle \varphi ^{2}\right\rangle _{0}$ is the renormalized VEV
in the boundary-free geometry for the conformal vacuum and the
sphere-induced contribution is given by
\begin{equation}
\left\langle \varphi ^{2}\right\rangle _{\mathrm{b}}=-\sum_{l=0}^{\infty }%
\frac{e^{-i\mu \pi }D_{l}}{S_{D}t^{D-1}}\int_{0}^{\infty }dx\,x\frac{\bar{Q}%
_{x-1/2}^{\mu }(u_{0})}{\bar{P}_{x-1/2}^{-\mu }(u_{0})}\frac{F_{\mu }^{(%
\mathrm{i})}(t,r,x)}{\sin (\pi x)}.  \label{phi2i}
\end{equation}%
Here
\begin{equation}
D_{l}=\frac{\left( 2l+n\right) \Gamma (l+n)}{\Gamma (n+1)l!}  \label{Dl}
\end{equation}%
is the degeneracy of the angular mode with given $l$ and we have introduced
the function
\begin{equation}
F_{\mu }^{(\mathrm{i})}(t,r,x)=J_{x}(mt)J_{-x}(mt)\frac{[P_{x-1/2}^{-\mu
}(u)]^{2}}{\sinh ^{D-2}r}.  \label{Fmu}
\end{equation}%
Note that in the absence of the spherical boundary the geometry under
consideration is homogeneous and the renormalized boundary-free VEV $%
\left\langle \varphi ^{2}\right\rangle _{0}$ will depend on the time
coordinate~only.

For a massless field one has
\begin{equation}
F_{\mu }^{(\mathrm{i})}(t,r,x)=\frac{\sin (\pi x)}{\pi x}\frac{%
[P_{x-1/2}^{-\mu }(u)]^{2}}{\sinh ^{D-2}r},  \label{Fmum0}
\end{equation}%
and the function (\ref{Fmu}) does not depend on the time coordinate. In this
case we get
\begin{equation}
\left\langle \varphi ^{2}\right\rangle _{\mathrm{b}}=\left( a/t\right)
^{D-1}\left\langle \varphi ^{2}\right\rangle _{\mathrm{b}}^{(\mathrm{st})},
\label{phi2m0}
\end{equation}%
where
\begin{equation}
\left\langle \varphi ^{2}\right\rangle _{\mathrm{b}}^{(\mathrm{st}%
)}=-\sum_{l=0}^{\infty }\frac{D_{l}e^{-i\mu \pi }}{\pi S_{D}a^{D-1}}%
\int_{0}^{\infty }dx\,\frac{\bar{Q}_{x-1/2}^{\mu }(u_{0})}{\bar{P}%
_{x-1/2}^{-\mu }(u_{0})}\frac{[P_{x-1/2}^{-\mu }(u)]^{2}}{\sinh ^{D-2}r},
\label{phi2st}
\end{equation}%
is the VEV for a massless conformally coupled field in a static negative
constant curvature space.

For a massive field and in the limit $t\rightarrow 0$ to the leading order
one finds%
\begin{equation}
\left\langle \varphi ^{2}\right\rangle _{\mathrm{b}}\approx \left(
a/t\right) ^{D-1}\left\langle \varphi ^{2}\right\rangle _{\mathrm{b}}^{(%
\mathrm{st})},\;t\rightarrow 0.  \label{phi2t0}
\end{equation}%
For large values of $t$, $mt\gg 1$, we use the asymptotic expression
\begin{equation}
J_{x}(mt)J_{-x}(mt)\approx \frac{\cos \left( \pi x\right) +\sin \left(
2mt\right) }{\pi mt}.  \label{Jlarge}
\end{equation}%
The leading term in the boundary-induced VEV is presented as%
\begin{equation}
\left\langle \varphi ^{2}\right\rangle _{\mathrm{b}}\approx -\frac{t^{-D}}{%
\pi S_{D}m}\sum_{l=0}^{\infty }D_{l}e^{-i\mu \pi }\int_{0}^{\infty }dx\,x%
\frac{\cos \left( \pi x\right) +\sin \left( 2mt\right) }{\sin (\pi x)}\frac{%
\bar{Q}_{x-1/2}^{\mu }(u_{0})}{\bar{P}_{x-1/2}^{-\mu }(u_{0})}\frac{%
[P_{x-1/2}^{-\mu }(u)]^{2}}{\sinh ^{D-2}r}.  \label{phi2larget}
\end{equation}%
As seen, for a massive field the late-time asymptotic of $\left\langle
\varphi ^{2}\right\rangle _{\mathrm{b}}$ contains two parts. The first one
is monotonically decreasing as $1/t^{D}$ and the behavior of the second one
is damping oscillatory, as $\sin \left( 2mt\right) /t^{D}$. For a massless
field the decay is monotonic, like $1/t^{D-1}$.

The VEV (\ref{phi2i}) diverges on the boundary $r=r_{0}$. For points near
the sphere the dominant contribution comes from large values of $l$ and $x$.
By taking into account that for large $x$ one has $J_{x}(mt)J_{-x}(mt)%
\approx \sin \left( \pi x\right) /(\pi x)$, we see that the leading term in
the asymptotic expansion near the sphere is obtained from that for (\ref%
{phi2st}) by the replacement $a\rightarrow t$:%
\begin{equation}
\left\langle \varphi ^{2}\right\rangle _{\mathrm{b}}\approx \frac{(1-2\delta
_{0B})\Gamma ((D-1)/2)}{(4\pi )^{(D+1)/2}[t(r_{0}-r)]^{D-1}}.
\label{phi2near}
\end{equation}%
Note that $t(r_{0}-r)$ is the proper distance from the sphere and the
leading term (\ref{phi2near}) coincides with that for a sphere in the
Minkowski bulk with the distance from the sphere replaced by the proper
distance in the geometry under consideration. As seen from (\ref{phi2near}),
near the sphere the boundary-induced contribution in the VEV of the field
squared is negative for Dirichlet boundary condition and positive for
non-Dirichlet boundary conditions. By taking into account that for small $r$
one has
\begin{equation}
\frac{\lbrack P_{x-1/2}^{-\mu }(u)]^{2}}{\sinh ^{D-2}r}\approx \frac{%
2^{-2\mu }r^{2l}}{\Gamma ^{2}(\mu +1)},  \label{smr}
\end{equation}%
we can see that at the sphere center the only nonzero contribution comes
from the mode with $l=0$ and we get%
\begin{equation}
\left. \left\langle \varphi ^{2}\right\rangle _{\mathrm{b}}\right\vert
_{r=0}=-\frac{\left( 2t\right) ^{1-D}e^{i(1-D/2)\pi }}{\pi ^{D/2}\Gamma (D/2)%
}\int_{0}^{\infty }dx\,x\frac{J_{x}(mt)J_{-x}(mt)}{\sin (\pi x)}\frac{\bar{Q}%
_{x-1/2}^{D/2-1}(u_{0})}{\bar{P}_{x-1/2}^{1-D/2}(u_{0})}.  \label{phi2r0}
\end{equation}%
For $D=3$ this expression is further simplified to%
\begin{equation}
\left. \left\langle \varphi ^{2}\right\rangle _{\mathrm{b}}\right\vert
_{r=0}=-\frac{1}{2\pi t^{2}}\int_{0}^{\infty }dx\,\frac{%
x^{2}J_{x}(mt)J_{-x}(mt)}{\sin (\pi x)\left( \frac{\beta +u_{0}-x}{\beta
+u_{0}+x}e^{2xr_{0}}-1\right) },  \label{phi2r0D3}
\end{equation}%
where $\beta =A/B$.

The integral in (\ref{phi2i}) is understood in the sense of the principal
value. For numerical calculations it is convenient to present it in an
alternative form, where the integrand has no poles. Let us consider the
integral of the form $\int_{0}^{\infty }dx\,f(x)/\sin (\pi x)$, understood
in the sense of the principal value. By using the formula (prime on the
summation sign means that the term $k=0$ should be taken with coefficient
1/2)%
\begin{equation}
\frac{1}{\sin (\pi x)}=\frac{2}{\pi }x\sideset{}{'}{\sum}_{k=0}^{\infty }%
\frac{(-1)^{k}}{x^{2}-k^{2}},  \label{sinsum}
\end{equation}%
and the fact that the principal value of the integral $\int_{0}^{\infty
}du/(u^{2}-k^{2})$ is zero, the integral is presented as%
\begin{equation}
\int_{0}^{\infty }dx\,\frac{f(x)}{\sin (\pi x)}=\frac{2}{\pi }%
\sideset{}{'}{\sum}_{k=0}^{\infty }(-1)^{k}\int_{0}^{\infty }dx\,\frac{%
xf(x)-kf(k)}{x^{2}-k^{2}}.  \label{intf3}
\end{equation}%
In this form the integrand is regular at the points $x=k$.

\begin{figure}[tbph]
\begin{center}
\begin{tabular}{cc}
\epsfig{figure=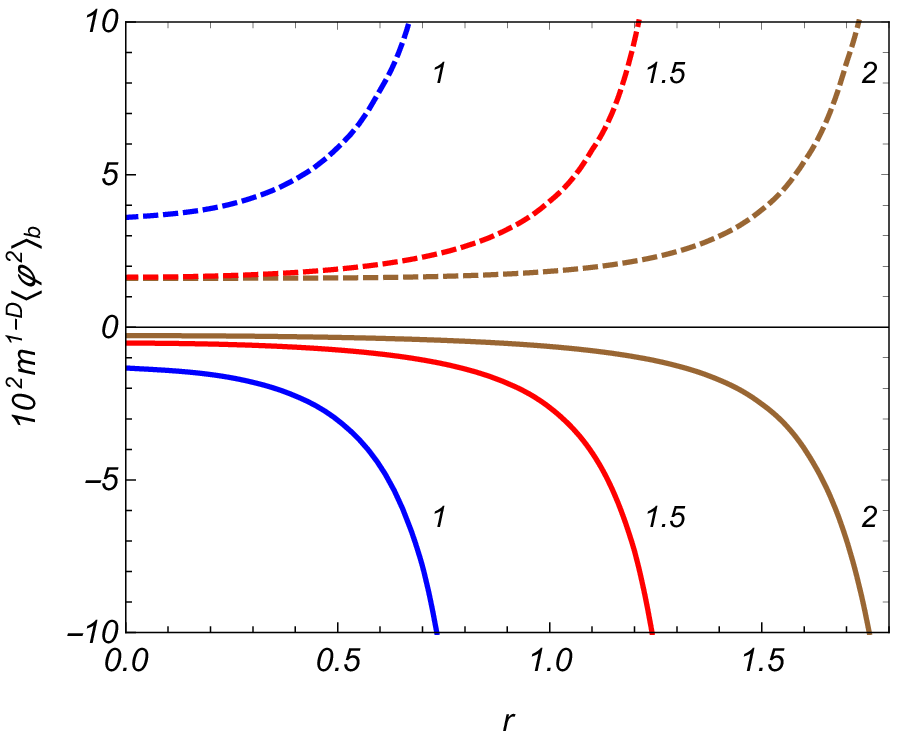,width=7.5cm,height=6.cm} & \quad %
\epsfig{figure=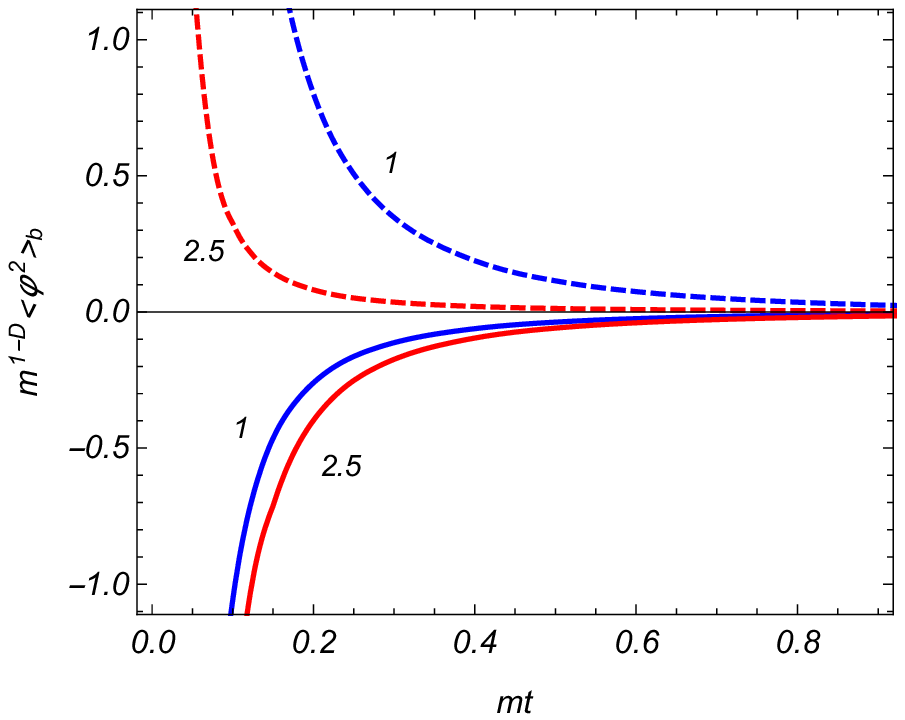,width=7.5cm,height=6.cm}%
\end{tabular}%
\end{center}
\caption{The sphere-induced VEV of the field squared for $D=3$ scalar field
as a function of the radial (left panel) and time (right panel) coordinates.
The left panel is plotted for $mt=1$ and the numbers near the curves
correspond to the values of the sphere radius $r_{0}$. For the right panel $%
r_{0}=2$ and the numbers near the curves present the values of the radial
coordinate. The full and dashed curves correspond to Dirichlet and Robin
(with $\protect\beta =-0.6$) boundary conditions, respectively.}
\label{fig2}
\end{figure}

In the left panel of Figure \ref{fig2}, for the Milne universe with $D=3$,
we have plotted the sphere-induced contribution in the VEV of the field
squared inside a spherical shell versus the radial coordinate $r$ for the
values of the sphere radius $r_{0}=1,1.5,2$ (numbers near the curves) and
for $mt=1$. The right panel presents the time-dependence of the same
quantity for fixed $r$ (numbers near the curves) and for $r_{0}=2$. The VEV
for the exterior region is displayed as well ($r=2.5$). The corresponding
analysis will be given in Section \ref{sec:Ext} below. On both the panels in
Figure \ref{fig2} the full curves correspond to Dirichlet boundary condition
and the dashed curves correspond to Robin boundary condition with $\beta
=-0.6$. Near the sphere the asymptotics are given by (\ref{phi2near}) and
the boundary-induced VEV, as a function of the radial coordinate,\ behaves
as $1/(r_{0}-r)^{2}$. For $mt\ll 1$ we have the conformal relation (\ref%
{phi2t0}) and the VEV behaves as $1/t^{2}$. In that region the
boundary-induced VEV\ in the field squared is a monotonic function of $mt$.
That is not the case for $mt>1$. As it has been shown before by the
asymptotic analysis (see (\ref{phi2larget})), for $mt\gg 1$ the VEV\
exhibits oscillatory damping behaviour and it can be either positive or
negative. That behavior is also confirmed by numerical analysis. For
example, in the case of Dirichlet boundary condition and for $r_{0}=2$, $r=1$
(the case corresponding to the right panel of Figure \ref{fig2}) with
increasing $mt$ the VEV $\left\langle \varphi ^{2}\right\rangle _{\mathrm{b}%
} $ becomes zero at $mt\approx 2.1$ and then takes the local maximum $%
\left\langle \varphi ^{2}\right\rangle _{\mathrm{b}}\approx 9.2\times
10^{-5}m^{2}$ at $mt\approx 2.4$. After that $\left\langle \varphi
^{2}\right\rangle _{\mathrm{b}}$ again becomes zero at $mt\approx 3.1$ with
the further damping oscillatory behavior as a function of $mt$.

\begin{figure}[tbph]
\begin{center}
\epsfig{figure=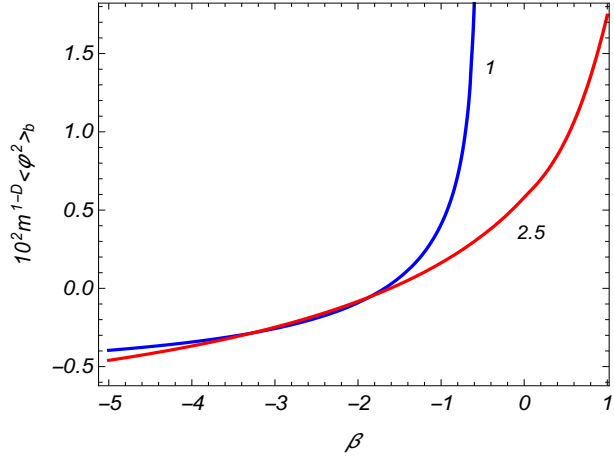,width=8cm,height=6cm}
\end{center}
\caption{The sphere-induced contribution in the VEV of the field squared for
$D=3$ scalar field versus the Robin coefficient. The graphs are plotted for $%
mt=1$, $r_{0}=2$ and the numbers near the curves correspond to the values of
the radial coordinate $r$.}
\label{fig3}
\end{figure}

Figure \ref{fig3} displays the dependence of the sphere-induced VEV on
the coefficient $\beta $ in Robin boundary condition for $D=3$, $mt=1$, $%
r_{0}=2$. The numbers near the curves correspond to the values of the radial
coordinate $r$. For the interior region we have taken $r=1$. As seen,
depending on the values of the Robin coefficient, the boundary-induced VEV
changes the sign. It becomes zero at $\beta \approx -1.65$. For $-\beta \gg
1 $ ($\beta |=-\infty $ corresponds to Dirichlet boundary condition) the VEV
is negative and it becomes positive with increasing $\beta $. For the
interior region (the graph for $r=1$), the~VEV increases when $\beta $
approaches the critical value $\beta _{0}^{(\mathrm{i})}(u_{0})\approx 0.537$
(see Table \ref{tab1}). For $\beta $ close to that critical value the dominant
contribution to the VEV (\ref{phi2i}) comes from the mode $l=0$. In the
range $\beta <-1.65$ the VEV $\langle \varphi ^{2}\rangle _{b}$ is negative
at $r=1$. By taking into account that near the sphere it is positive, we~conclude that for values of $\beta $ in that range the VEV\ $\langle \varphi
^{2}\rangle _{b}$, considered as a function of $r$, changes the~sign.

In the evaluation of the VEV of the energy-momentum tensor the covariant
d'Alembertian of the VEV of the field squared is required. The
sphere-induced contribution in the latter is presented in the~form%
\begin{eqnarray}
\nabla _{p}\nabla ^{p}\langle \varphi ^{2}\rangle _{b} &=&\sum_{l=0}^{\infty
}\frac{e^{-i\mu \pi }D_{l}}{S_{D}t^{D+1}}\int_{0}^{\infty }dx\,\frac{x}{\sin
(x\pi )}\frac{\bar{Q}_{x-1/2}^{\mu }(u_{0})}{\bar{P}_{x-1/2}^{-\mu }(u_{0})}
\notag \\
&&\times \left[ (u^{2}-1)\partial _{u}^{2}+Du\partial _{u}-t^{2}\partial
_{t}^{2}+\left( D-2\right) t\partial _{t}\right] F_{\mu }^{(\mathrm{i}%
)}(t,r,x).  \label{DalPhi2}
\end{eqnarray}%
Note that for the radial part of the function $F_{\mu }^{(\mathrm{i}%
)}(t,r,x) $ we have the equation
\begin{equation}
\left[ (u^{2}-1)\partial _{u}^{2}+Du\partial _{u}-x^{2}+\frac{nD+1}{4}-\frac{%
l(l+n)}{u^{2}-1}\right] \frac{P_{x-1/2}^{-\mu }(u)}{(u^{2}-1)^{\frac{D-2}{4}}%
}=0.  \label{EqP}
\end{equation}%
This equation can be used to exclude the second derivative $\partial
_{u}^{2}F_{\mu }^{(\mathrm{i})}(t,r,x)$ from the expressions of the VEV for
the energy-momentum tensor.

\subsection{VEV of the Energy-Momentum Tensor}

Another important local characteristic for the vacuum state is the VEV of
the energy-momentum tensor. Having the Hadamard function and the VEV of the
field squared, that VEV is evaluated by using the formula
\begin{equation}
\langle T_{ik}\rangle =\frac{1}{2}\lim_{x^{\prime }\rightarrow x}\partial
_{i^{\prime }}\partial _{k}G(x,x^{\prime })+\left[ \left( \xi -1/4\right)
g_{ik}\nabla _{p}\nabla ^{p}-\xi \nabla _{i}\nabla _{k}-\xi \mathcal{R}_{ik}%
\right] \langle \varphi ^{2}\rangle ,  \label{EMT}
\end{equation}%
where for the geometry at hand the Ricci tensor vanishes, $\mathcal{R}%
_{ik}=0 $, and for the d'Alembertian acted on the VEV of the field squared
one has (\ref{DalPhi2}). In (\ref{EMT}) we have used the expression for the
classical energy-momentum tensor that differs from the standard one (given,
for example, in \cite{Birr82B}) by the term that vanishes on the solutions
of the field equation (see \cite{Saha04c}). The VEV\ of the energy-momentum
tensor is decomposed as%
\begin{equation}
\langle T_{ik}\rangle =\langle T_{ik}\rangle _{0}+\langle T_{ik}\rangle _{%
\mathrm{b}},  \label{Tik0s}
\end{equation}%
with the boundary-free and the sphere-induced contributions $\langle
T_{ik}\rangle _{0}$ and $\langle T_{ik}\rangle _{\mathrm{b}}$, respectively.
Similar to the case of the field squared, the renormalized boundary-free
contribution $\langle T_{i}^{k}\rangle _{0}$ will depend on the time
coordinate only. It is diagonal and the corresponding vacuum stresses are
isotropic $\langle T_{1}^{1}\rangle _{0}=\langle T_{2}^{2}\rangle
_{0}=\cdots =\langle T_{D}^{D}\rangle _{0}$. The background geometry is flat
and for a conformally coupled massless field the vacuum energy-momentum
tensor is traceless, $\langle T_{k}^{k}\rangle _{0}=0$ (the conformal
anomaly is absent). In this case $\langle T_{0}^{0}\rangle _{0}=-D\langle
T_{1}^{1}\rangle _{0}$.

The diagonal components of the sphere-induced VEV of the energy-momentum
tensor are presented in the form (no summation over $k$)%
\begin{equation}
\langle T_{k}^{k}\rangle _{\mathrm{b}}=\sum_{l=0}^{\infty }\frac{e^{-i\mu
\pi }D_{l}}{S_{D}t^{D+1}}\int_{0}^{\infty }dx\,\frac{x}{\sin (x\pi )}\frac{%
\bar{Q}_{x-1/2}^{\mu }(u_{0})}{\bar{P}_{x-1/2}^{-\mu }(u_{0})}\hat{F}%
_{(k)}F_{\mu }^{(\mathrm{i})}(t,r,x),  \label{Tkk}
\end{equation}%
where the operators for separate components are given by the expressions
\begin{eqnarray}
\hat{F}_{(0)} &=&-\frac{1}{4}t^{2}\partial _{t}^{2}-\left( D\xi -\frac{D-2}{4%
}\right) t\partial _{t}-m^{2}t^{2}+x^{2}  \notag \\
&&+\left( \xi -\frac{1}{4}\right) \left[ (u^{2}-1)\partial
_{u}^{2}+Du\partial _{u}\right] +D\left( D-1\right) \left( \xi -\xi
_{D}\right) ,  \notag \\
\hat{F}_{(1)} &=&\left( \frac{1}{4}-\xi \right) t^{2}\partial _{t}^{2}+\left[
\left( D-1\right) \xi -\frac{D-2}{4}\right] t\partial _{t}-\left( D-1\right)
\left( \xi -\frac{D-1}{4}\right)  \notag \\
&&+\frac{1}{4}\left( u^{2}-1\right) \partial _{u}^{2}+\left[ \left(
D-1\right) \xi +\frac{D}{4}\right] u\partial _{u}-\frac{l(l+D-2)}{u^{2}-1}%
-x^{2},  \notag \\
\hat{F}_{(2)} &=&\left( \frac{1}{4}-\xi \right) t^{2}\partial _{t}^{2}+\left[
\left( D-1\right) \xi -\frac{D-2}{4}\right] t\partial _{t}+\frac{l(l+D-2)}{%
(D-1)\left( u^{2}-1\right) }  \notag \\
&&+\left( \xi -\frac{1}{4}\right) \left( u^{2}-1\right) \partial _{u}^{2}+%
\left[ \left( D-1\right) \xi -\frac{D}{4}\right] u\partial _{u}-\left(
D-1\right) \xi ,  \label{F(0)}
\end{eqnarray}%
and $\hat{F}_{(k)}=\hat{F}_{(2)}$ for $k=3,\ldots ,D$. In (\ref{F(0)}), $\xi
_{D}=(D-1)/(4D)$ is the curvature coupling parameter for a conformally
coupled scalar field. The boundary-induced VEVs (\ref{Tkk}) obey the trace
relation%
\begin{equation}
\langle T_{k}^{k}\rangle _{\mathrm{b}}=\left[ D(\xi -\xi _{D})\nabla
_{p}\nabla ^{p}+m^{2}\right] \langle \varphi ^{2}\rangle _{\mathrm{b}}.
\label{tracerel}
\end{equation}%
For a conformally coupled massless field the boundary-induced VEV of the
energy-momentum tensor is traceless. The background geometry is flat and the
latter is the case for the boundary-free part $\langle T_{ik}\rangle _{0}$
as well.

The problem under consideration is inhomogeneous with respect to the
coordinates $t$ and $r$. As a consequence of that the VEV\ of the
energy-momentum tensor has nonzero off-diagonal component%
\begin{eqnarray}
\langle T_{0}^{1}\rangle &=&\sum_{l=0}^{\infty }\frac{e^{-i\mu \pi }D_{l}}{%
4S_{D}t^{D+2}}\int_{0}^{\infty }dx\,\frac{x}{\sin (x\pi )}\frac{\bar{Q}%
_{x-1/2}^{\mu }(u_{0})}{\bar{P}_{x-1/2}^{-\mu }(u_{0})}  \notag \\
&&\times \left[ 4D\left( \xi -\xi _{D}\right) +\left( 1-4\xi \right)
t\partial _{t}\right] \partial _{r}F_{\mu }^{(\mathrm{i})}(t,r,x).
\label{T01fl}
\end{eqnarray}%
This component describes energy flux along the radial direction. The
boundary-free part in $\langle T_{0}^{1}\rangle $ vanishes and (\ref{T01fl})
is induced by the sphere. For a massless field the function $F_{\mu }^{(%
\mathrm{i})}(t,r,x)$ does not depend on time and in the case of conformal
coupling ($\xi =\xi _{D}$) the energy flux vanishes. Of~course, we could
expect this result on the base of the conformal relation of the problem at
hand to the corresponding problem in static spacetime with negative constant
curvature space.

The sphere-induced VEVs obey the covariant conservation equation for the
energy-momentum tensor, $\nabla _{k}\langle T_{i}^{k}\rangle _{\mathrm{b}}=0$
with $\langle T_{0}^{1}\rangle _{\mathrm{b}}=\langle T_{0}^{1}\rangle $. The
latter is reduced to the following two relations
\begin{eqnarray}
\frac{\partial _{t}\left( t^{D}\langle T_{0}^{0}\rangle _{\mathrm{b}}\right)
}{t^{D}}+\frac{\partial _{r}\left( \sinh ^{D-1}r\langle T_{0}^{1}\rangle
\right) }{\sinh ^{D-1}r}-\frac{1}{t}\langle T_{1}^{1}\rangle _{\mathrm{b}}-%
\frac{D-1}{t}\langle T_{2}^{2}\rangle _{\mathrm{b}} &=&0,  \notag \\
\frac{\partial _{t}\left( t^{D}\langle T_{1}^{0}\rangle \right) }{t^{D}}+%
\frac{\partial _{r}\left( \sinh ^{D-1}r\langle T_{1}^{1}\rangle _{\mathrm{b}%
}\right) }{\sinh ^{D-1}r}-\frac{D-1}{\tanh r}\langle T_{2}^{2}\rangle _{%
\mathrm{b}} &=&0.  \label{Contequ}
\end{eqnarray}%
The component $\langle T_{0}^{0}\rangle _{\mathrm{b}}$ determines the
contribution of the spherical boundary to the vacuum energy density. The
boundary-induced part of the vacuum energy in the spatial volume $V$ with a
boundary $\partial V$ is given by (in the coordinates with $g_{00}=1$)
\begin{equation}
E_{V}^{(\mathrm{b})}=\int_{V}d^{D}x\sqrt{|g|}\langle T_{0}^{0}\rangle _{%
\mathrm{b}},  \label{EbV}
\end{equation}%
where $g$ is the determinant of the metric tensor. From the equation $\nabla
_{k}\langle T_{0}^{k}\rangle _{\mathrm{b}}=0$ (the first equation in (\ref%
{Contequ})) we get
\begin{equation}
\partial _{t}E_{V}^{(\mathrm{b})}=-\int_{\partial V}d^{D-1}x\,\sqrt{h}%
n_{\alpha }\langle T_{0}^{\alpha }\rangle +\frac{1}{t}\int_{V}d^{D}x\sqrt{|g|%
}\langle T_{\alpha }^{\alpha }\rangle _{\mathrm{b}},  \label{EnCons}
\end{equation}%
where the Greek indices run over $1,2,\ldots ,D$. In (\ref{EnCons}), $%
n_{\alpha }$, $g^{\alpha \beta }n_{\alpha }n_{\beta }=-1$, is the external
normal to the boundary $\partial V$ and $h$ is the determinant of the
induced spatial metric $h_{\alpha \beta }=-g_{\alpha \beta }-n_{\alpha
}n_{\beta }$ on $\partial V$. From~(\ref{EnCons}) we see that the energy
flux density per unit proper surface area is given by $n_{\alpha }\langle
T_{0}^{\alpha }\rangle $. Note that for a spherical boundary $\partial V$ of
radius $r$ with the center at $r=0$ one has $n_{\alpha }=\pm \delta _{\alpha
}^{1}t$, where the upper/lower sign corresponds to the volume $V$
inside/outside the sphere. The energy flux is directed from the sphere if $%
n_{\alpha }\langle T_{0}^{\alpha }\rangle <0$ and towards the sphere if $%
n_{\alpha }\langle T_{0}^{\alpha }\rangle >0$. From here it follows that for
the interior region the energy flux is directed from the sphere for $\langle
T_{0}^{1}\rangle <0$ and towards the sphere for $\langle T_{0}^{1}\rangle >0$%
.

For a massless field the function $F_{\mu }^{(\mathrm{i})}(t,r,x)$ is given
by (\ref{Fmum0}) and does not depend on $t$. In this case the terms in the
expressions (\ref{F(0)}) for the operators $\hat{F}_{(k)}$ containing
derivatives over $t$ can be removed. For a conformally coupled massless
field one gets%
\begin{eqnarray}
\hat{F}_{(0)} &=&x^{2}-\frac{1}{4D}\left[ (u^{2}-1)\partial
_{u}^{2}+Du\partial _{u}\right] ,  \notag \\
\hat{F}_{(1)} &=&\frac{1}{4}\left( u^{2}-1\right) \partial _{u}^{2}+\frac{%
2D^{2}-2D+1}{4D}u\partial _{u}+\frac{\left( D-1\right) ^{3}}{4D}-\frac{%
l(l+D-2)}{u^{2}-1}-x^{2},  \notag \\
\hat{F}_{(2)} &=&-\frac{1}{4D}\left[ \left( u^{2}-1\right) \partial
_{u}^{2}+\left( 2D-1\right) u\partial _{u}\right] +\frac{l(l+D-2)}{%
(D-1)\left( u^{2}-1\right) }-\frac{\left( D-1\right) ^{2}}{4D},
\label{F(0)c}
\end{eqnarray}%
and the off-diagonal component of the vacuum energy-momentum tensor
vanishes. Now we have the conformal relation (no summation over $k$) $%
\langle T_{k}^{k}\rangle _{\mathrm{b}}=\left( a/t\right) ^{D+1}\langle
T_{k}^{k}\rangle _{\mathrm{b}}^{(\mathrm{st})}$, where the VEV in static
spacetime with constant negative curvature space is given by
\begin{equation}
\langle T_{k}^{k}\rangle _{\mathrm{b}}^{(\mathrm{st})}=\sum_{l=0}^{\infty }%
\frac{e^{-i\mu \pi }D_{l}}{\pi S_{D}a^{D+1}}\int_{0}^{\infty }dx\,\frac{\bar{%
Q}_{x-1/2}^{\mu }(u_{0})}{\bar{P}_{x-1/2}^{-\mu }(u_{0})}\hat{F}_{(k)}\frac{%
[P_{x-1/2}^{-\mu }(u)]^{2}}{\sinh ^{D-2}r},  \label{Tkkst}
\end{equation}%
with $\hat{F}_{(k)}$ from (\ref{F(0)c}). It can be checked that (\ref{Tkkst}%
) coincides with the result from \cite{Bell14} for a conformally coupled
massless field.

The general expressions for the VEVs of the components of the
energy-momentum tensor are rather complicated and we turn to the
investigation of their behavior in the asymptotic regions of the parameters.
In the limit $t\rightarrow 0$ (early stages of the expansion) one has $%
J_{x}(mt)J_{-x}(mt)\rightarrow \sin (\pi x)/(\pi x)$ and to the leading
order for the diagonal components we get $\langle T_{k}^{k}\rangle _{\mathrm{%
b}}\approx \left( a/t\right) ^{D+1}\langle T_{k}^{k}\rangle _{\mathrm{b}}^{(%
\mathrm{st})}$, where $\langle T_{k}^{k}\rangle _{\mathrm{b}}^{(\mathrm{st}%
)} $ is the corresponding VEV for a massless field with curvature coupling
parameter $\xi $ in static spacetime with a negative constant curvature
space. For the energy flux and for non-conformally coupled fields ($\xi \neq
\xi _{D}$) we have
\begin{equation}
\langle T_{0}^{1}\rangle \approx \frac{D\left( \xi -\xi _{D}\right) }{\pi
S_{D}t^{D+2}}\sum_{l=0}^{\infty }e^{-i\mu \pi }D_{l}\int_{0}^{\infty }dx\,%
\frac{\bar{Q}_{x-1/2}^{\mu }(u_{0})}{\bar{P}_{x-1/2}^{-\mu }(u_{0})}\partial
_{r}\frac{[P_{x-1/2}^{-\mu }(u)]^{2}}{\sinh ^{D-2}r}.  \label{T01sm}
\end{equation}%
The leading term in the right-hand side does not depend on the field mass.
In the case of massless fields the relation (\ref{T01sm}) is exact. For a
conformally coupled field we need to keep the next to the leading order
terms in the expansion of the product $J_{x}(mt)J_{-x}(mt)$. This leads to
the result%
\begin{equation}
\langle T_{0}^{1}\rangle \approx -\frac{m^{2}t^{-D}}{4\pi DS_{D}}%
\sum_{l=0}^{\infty }e^{-i\mu \pi }D_{l}\int_{0}^{\infty }dx\,\frac{\bar{Q}%
_{x-1/2}^{\mu }(u_{0})}{\bar{P}_{x-1/2}^{-\mu }(u_{0})}\frac{1}{1-x^{2}}%
\partial _{r}\frac{[P_{x-1/2}^{-\mu }(u)]^{2}}{\sinh ^{D-2}r},
\label{T01sm2}
\end{equation}%
for $\xi =\xi _{D}$. Here the integrand has a simple pole at $x=1$ and the
integral is understood in the sense of the principal value.

Now let us consider the late stages of the expansion, $mt\gg 1$. By using
the corresponding asymptotic expression (\ref{Jlarge}), we see that the
dominant contributions come from the term with $\sin \left( 2mt\right) $ in (%
\ref{Jlarge}) and from the parts in (\ref{F(0)}) with $t^{2}\partial
_{t}^{2} $. The leading term in the VEV\ of the energy density vanishes. To
the leading order, the vacuum stresses are isotropic and (no summation over $%
k=1,2,\ldots ,D$)
\begin{equation}
\langle T_{k}^{k}\rangle _{\mathrm{b}}\approx \frac{4\xi -1}{\pi S_{D}t^{D}}%
m\sin \left( 2mt\right) \sum_{l=0}^{\infty }e^{-i\mu \pi
}D_{l}\int_{0}^{\infty }dx\,\frac{x}{\sin (\pi x)}\frac{\bar{Q}_{x-1/2}^{\mu
}(u_{0})}{\bar{P}_{x-1/2}^{-\mu }(u_{0})}\frac{[P_{x-1/2}^{-\mu }(u)]^{2}}{%
\sinh ^{D-2}r}.  \label{Tkklarge}
\end{equation}%
For the energy flux from (\ref{T01fl}) we get%
\begin{equation}
\langle T_{0}^{1}\rangle \approx \frac{\left( 1-4\xi \right) \cos \left(
2mt\right) }{2\pi S_{D}t^{D+2}}\sum_{l=0}^{\infty }e^{-i\mu \pi
}D_{l}\int_{0}^{\infty }dx\frac{x}{\sin (x\pi )}\frac{\bar{Q}_{x-1/2}^{\mu
}(u_{0})}{\bar{P}_{x-1/2}^{-\mu }(u_{0})}\partial _{r}\frac{[P_{x-1/2}^{-\mu
}(u)]^{2}}{\sinh ^{D-2}r}.  \label{T01large}
\end{equation}%
The leading term in the energy density can be found from the first relation
in (\ref{Contequ}):%
\begin{equation}
\langle T_{0}^{0}\rangle _{\mathrm{b}}\approx -\frac{D\cot \left( 2mt\right)
}{2mt}\langle T_{1}^{1}\rangle _{\mathrm{b}},  \label{T00large}
\end{equation}%
with $\langle T_{1}^{1}\rangle _{\mathrm{b}}$ from (\ref{Tkklarge}). Note
that the dependence on the curvature coupling parameter enters in the form
of the coefficient $4\xi -1$. In particular, the leading term in the
asymptotic for a minimally coupled field is obtained from that for a
conformally coupled field multiplying by $D$. As seen from the asymptotic
expressions (\ref{Tkklarge})--(\ref{T00large}), at late stages of the
expansion the behavior of the sphere-induced contributions in the VEVs, as
functions of the time coordinate, is oscillatory damping.

By using (\ref{smr}), we see that at the sphere center the nonzero
contributions come from the terms $l=0,1$. The energy flux vanishes at the
sphere center and for small $r$ it is given by
\begin{equation}
\langle T_{0}^{1}\rangle \approx \frac{\left( 4\pi \right) ^{-D/2}e^{-iD\pi
/2}r}{D\Gamma (D/2)t^{D+2}}\int_{0}^{\infty }dx\,x\frac{\bar{Q}%
_{x-1/2}^{D/2}(u_{0})}{\bar{P}_{x-1/2}^{-D/2}(u_{0})}\left[ 4D\left( \xi
-\xi _{D}\right) +\left( 1-4\xi \right) t\partial _{t}\right] \frac{%
J_{x}(mt)J_{-x}(mt)}{\sin (x\pi )}.  \label{T01r0}
\end{equation}%
Similar to the field squared, the VEV\ of the energy-momentum tensor is
divergent on the sphere. The leading terms in the energy density and
tangential stresses are found from (\ref{Tkk}) by using the asymptotic $%
J_{x}(mt)J_{-x}(mt)\approx \sin \left( \pi x\right) /(\pi x)$. They are
given by (no summation over $k=0,2,3,\ldots ,D$)
\begin{equation}
\langle T_{k}^{k}\rangle _{\mathrm{b}}\approx \frac{D\Gamma ((D+1)/2)(\xi
-\xi _{D})}{2^{D}\pi ^{(D+1)/2}[t(r_{0}-r)]^{D+1}}(2\delta _{0B}-1).
\label{Tkknear}
\end{equation}%
The leading term in the radial stress vanishes. In order to find the
corresponding asymptotic we first consider the leading term for the energy
flux. The latter is found from the first equation of (\ref{Contequ}):
\begin{equation}
\langle T_{0}^{1}\rangle \approx \frac{r_{0}-r}{t}\langle T_{0}^{0}\rangle _{%
\mathrm{b}}.  \label{T01near}
\end{equation}%
Combining this result with (\ref{Tkknear}), from the second relation of (\ref%
{Contequ}) we get
\begin{equation}
\langle T_{1}^{1}\rangle _{\mathrm{b}}\approx (r_{0}-r)\frac{1-1/D}{\tanh
r_{0}}\langle T_{0}^{0}\rangle _{\mathrm{b}}.  \label{T11near}
\end{equation}%
Note that the leading term (\ref{Tkknear}) in the energy density and the
tangential stresses coincides with that for spherical boundary in the
Minkowski bulk where the distance from the sphere is replaced by the proper
distance $t(r_{0}-r)$ for the geometry (\ref{ds2}). For a conformally
coupled field the leading terms in the asymptotic expansion over the
distance from the boundary vanish and the divergences on the sphere are
weaker. As seen from (\ref{Tkknear}), for a minimally coupled field the
energy density near the sphere is negative for Dirichlet boundary condition
and positive for non-Dirichlet conditions. Combining this with (\ref{T01near}%
) we see that for a minimally coupled field and near the sphere the energy
flux in the interior region is directed from the sphere for Dirichlet
boundary condition and towards the sphere for non-Dirichlet boundary
conditions.

\begin{figure}[tbph]
\begin{center}
\begin{tabular}{cc}
\epsfig{figure=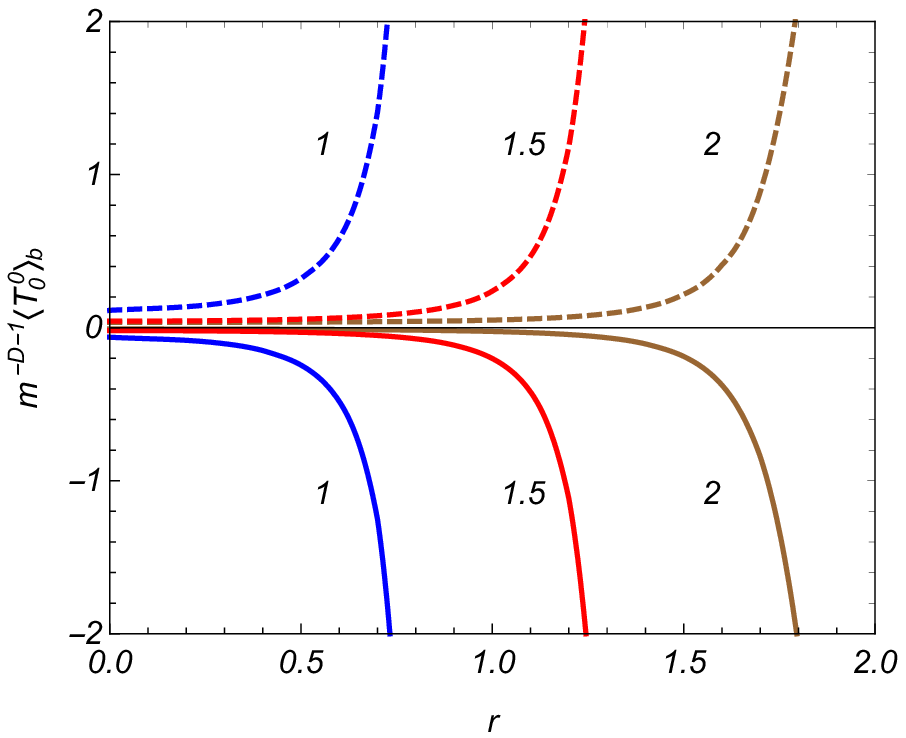,width=7.5cm,height=6.cm} & \quad %
\epsfig{figure=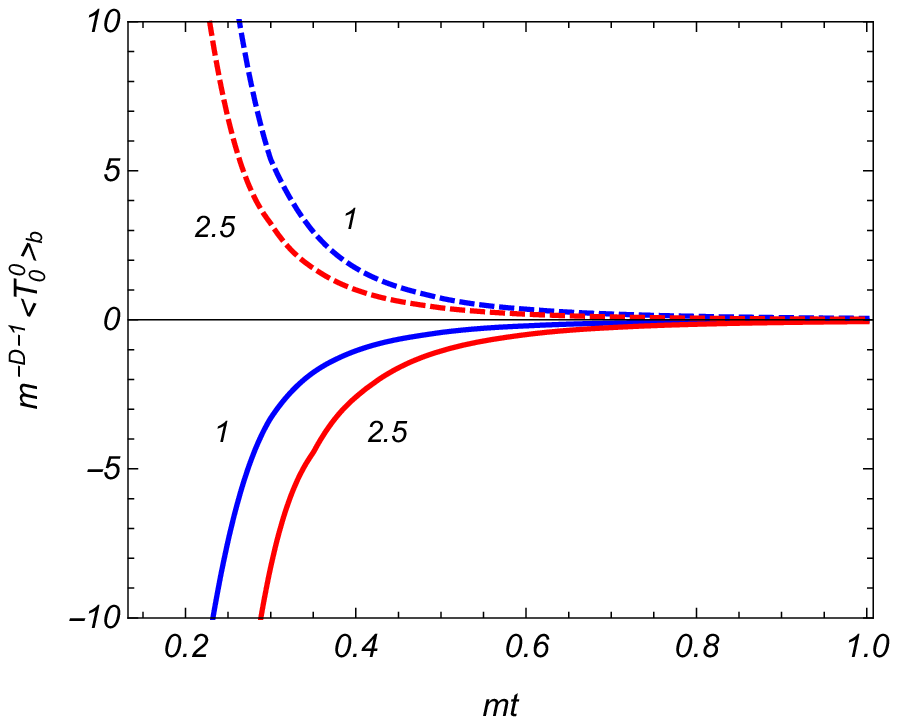,width=7.5cm,height=6.cm}%
\end{tabular}%
\end{center}
\caption{The same as in figure \protect\ref{fig2} for the boundary-induced
energy density in the case of a minimally coupled field.}
\label{fig4}
\end{figure}

In Figure \ref{fig4}, for the $D=3$ Milne universe and for a minimally
coupled field, the boundary-induced VEV in the energy density is displayed
as a function of the radial (left panel) and time (right panel) coordinates.
The full and dashed curves correspond to Dirichlet and Robin boundary
conditions. For the latter we have taken $\beta =-0.6$. For the left panel $%
mt=1$ and the numbers near the curves correspond to the values of the sphere
radius. The graphs on the right panel are plotted for $r_{0}=2$ and the
numbers near the curves are the values of the radial coordinate $r$. Near
the sphere the energy density behaves as $1/(r_{0}-r)^{4}$. It is negative
for Dirichlet boundary condition and positive for non-Dirichlet conditions.
For an example presented in the left panel of Figure \ref{fig4} in the case
of Robin boundary condition the energy density is positive for all values of
the radial coordinate inside the sphere. That is not the case for general
values of $\beta $. Depending on the latter, the energy density may change
the sign. Considered as a function of the time coordinate, in the initial
stages of the expansion the sphere-induced energy density behaves like $%
1/t^{4}$. At late stages, $mt\gg 1$, it decays as $\cos (2mt)/t^{4}$. Recall
that for a massless field the late-time decay is monotonic, as $1/t^{4}$.

\begin{figure}[tbph]
\begin{center}
\begin{tabular}{cc}
\epsfig{figure=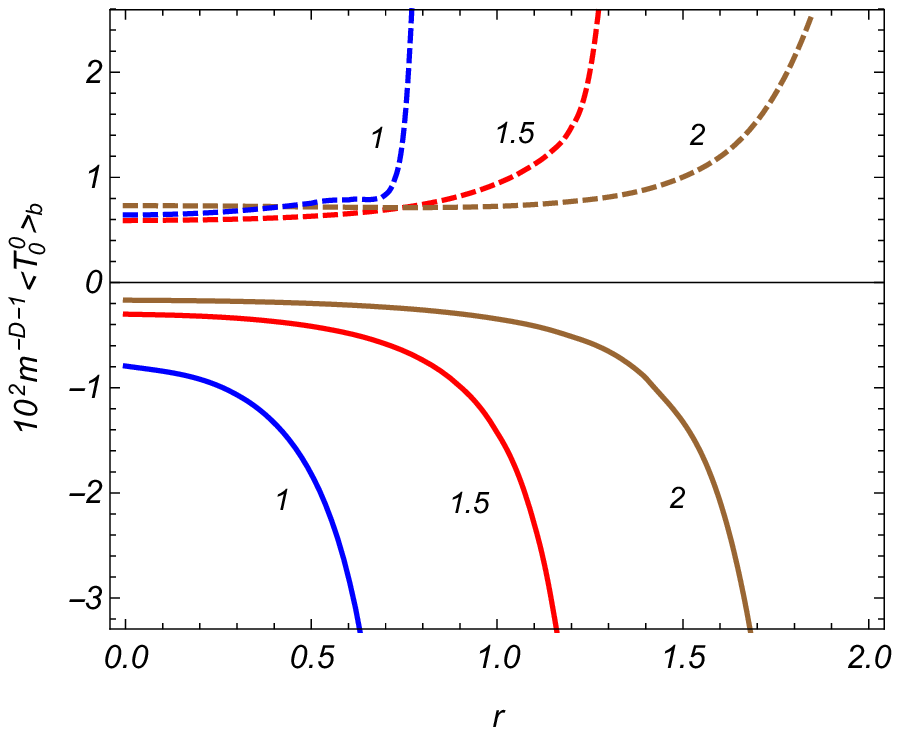,width=7.5cm,height=6.cm} & \quad %
\epsfig{figure=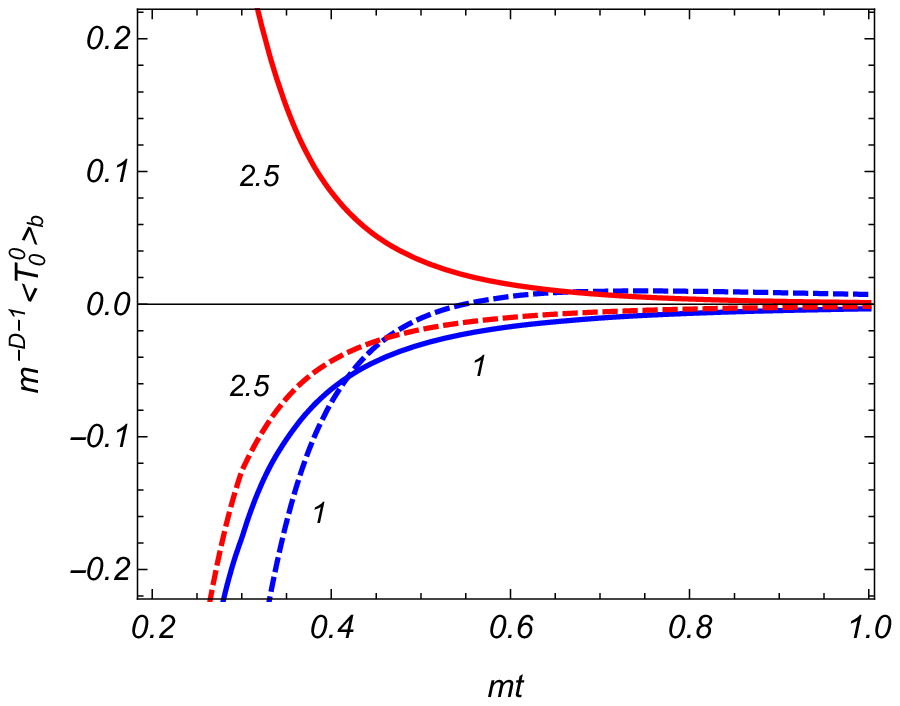,width=7.5cm,height=6.cm}%
\end{tabular}%
\end{center}
\caption{The same as in figure \protect\ref{fig4} for a conformally coupled
field.}
\label{fig5}
\end{figure}

The same graphs as in Figure \ref{fig4} are plotted for a conformally
coupled scalar field in Figure \ref{fig5}. In this case the divergence of
the boundary induced VEV on the sphere is weaker. At late stages,
corresponding to $mt\gg 1$, the leading term in the asymptotic expansion of $%
\langle T_{0}^{0}\rangle _{\mathrm{b}}$ for a conformally coupled field
differs from that for a minimally coupled field by an additional coefficient
$1/D$ (with~$D=3$ for the example in Figure \ref{fig5}).

\begin{figure}[tbph]
\begin{center}
\begin{tabular}{cc}
\epsfig{figure=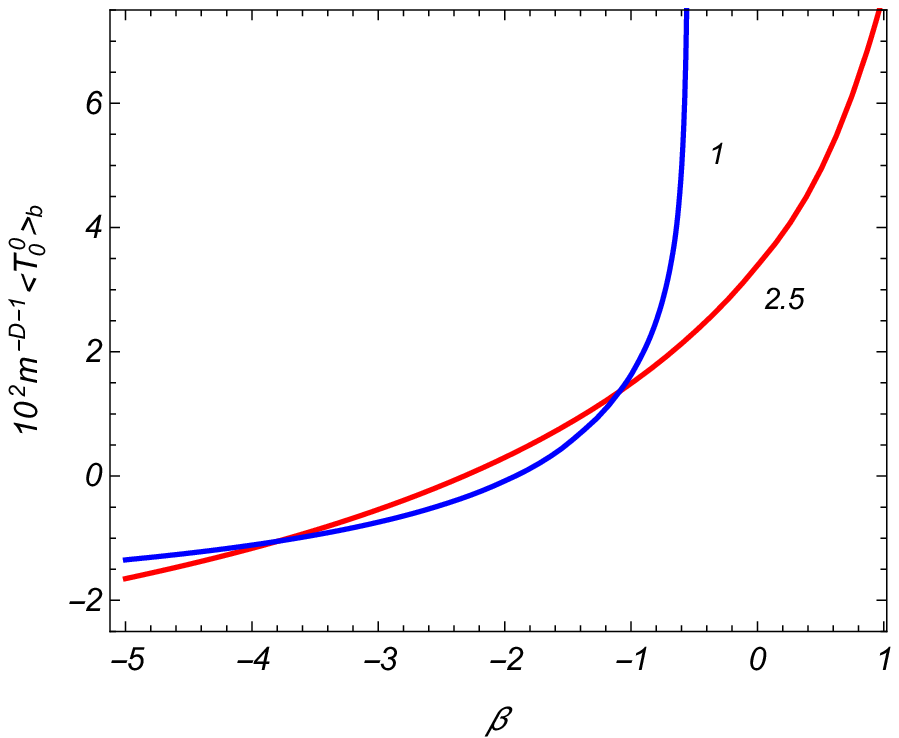,width=7.5cm,height=6.cm} & \quad %
\epsfig{figure=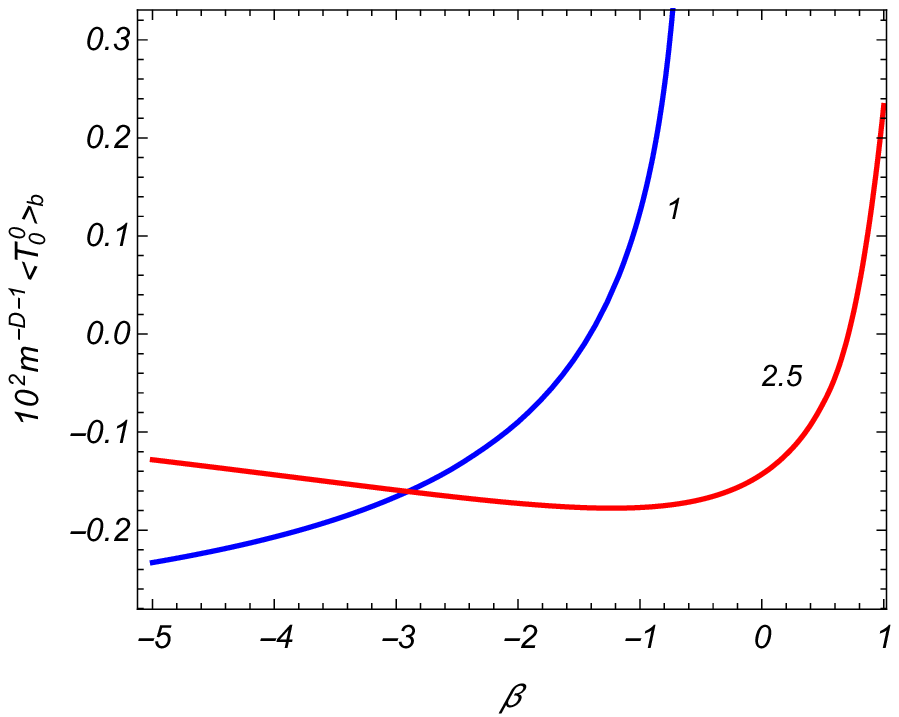,width=7.5cm,height=6.cm}%
\end{tabular}%
\end{center}
\caption{The same as in figure \protect\ref{fig3} for the boundary-induced
contribution in the VEV of the energy density for the cases of minimally
(left panel) and conformally (right panel) coupled fields.}
\label{fig6}
\end{figure}

Figure \ref{fig6} presents the dependence of the sphere-induced VEV in the
energy density on the coefficient $\beta $ in the Robin boundary condition
for minimally (left panel) and conformally (right panel) coupled $D=3$
scalar fields. The graphs are plotted for $mt=1$, $r_{0}=2$ and the numbers
near the curves are the values of the coordinate $r$. For the interior
region we have taken $r=1$ and the corresponding VEV becomes zero at $\beta
\approx -1.86$ for a minimally coupled field and at $\beta \approx -2.29$
for a conformally coupled field. For smaller values of $\beta $ the energy
density is negative. By taking into account that for a minimally coupled
field the corresponding energy density is positive near the sphere we see
that it changes the sign as a function of the radial coordinate for
sufficiently small values of $\beta $.

\begin{figure}[tbph]
\begin{center}
\begin{tabular}{cc}
\epsfig{figure=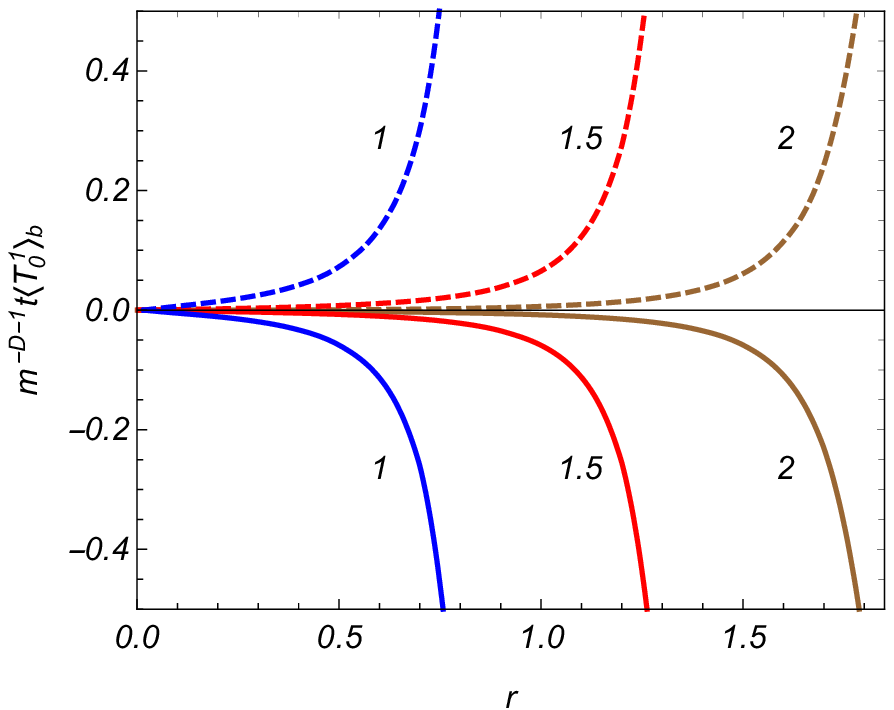,width=7.5cm,height=6.cm} & \quad %
\epsfig{figure=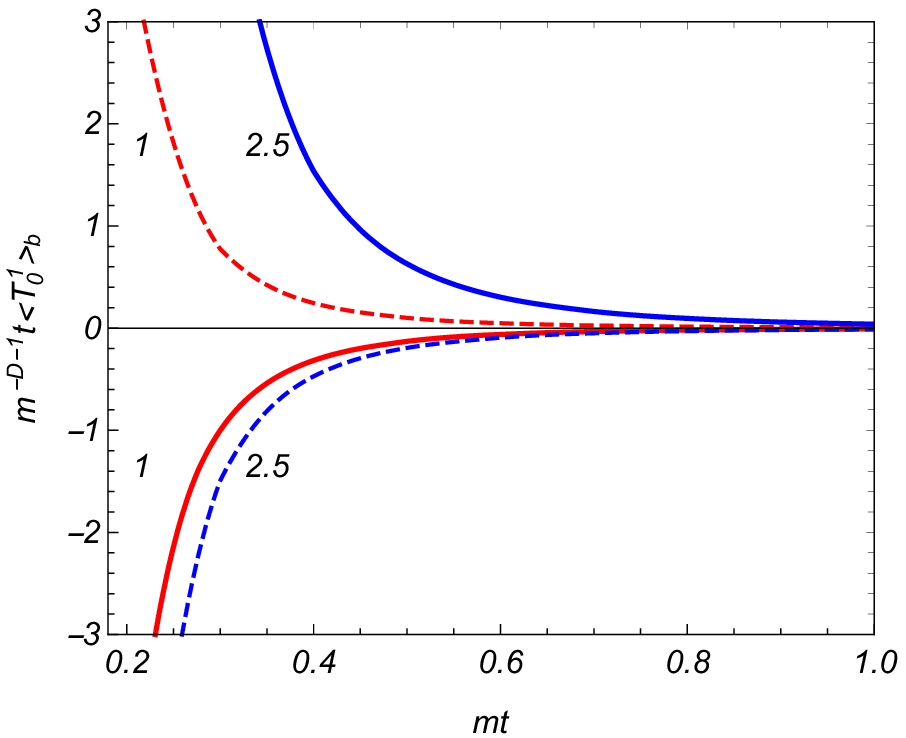,width=7.5cm,height=6.cm}%
\end{tabular}%
\end{center}
\caption{The same as in figure \protect\ref{fig4} for the energy flux
density.}
\label{fig7}
\end{figure}

\begin{figure}[tbph]
\begin{center}
\begin{tabular}{cc}
\epsfig{figure=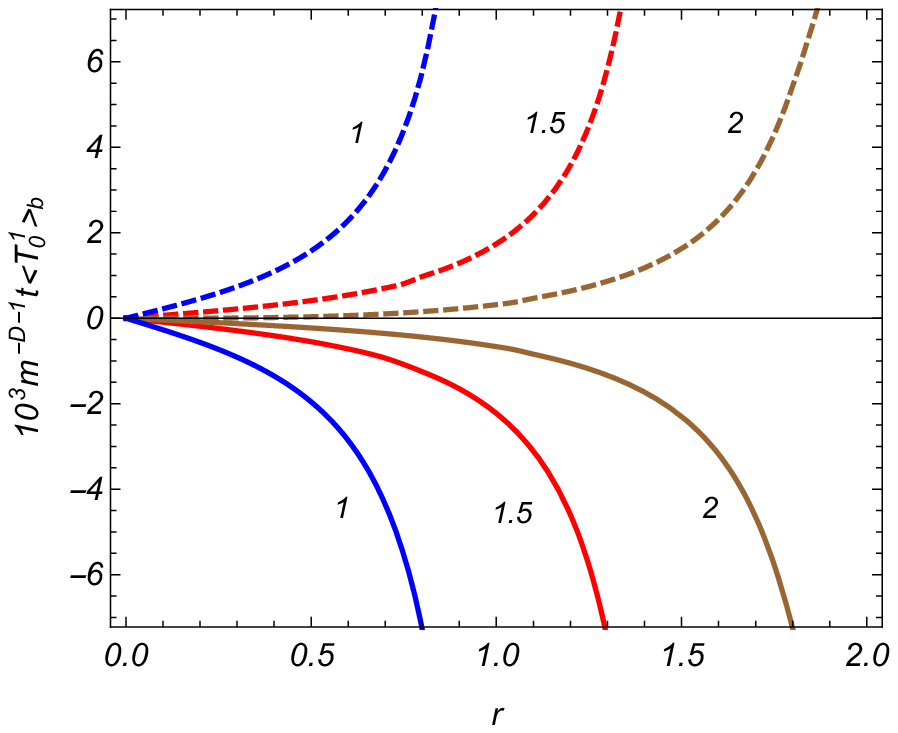,width=7.5cm,height=6.cm} & \quad %
\epsfig{figure=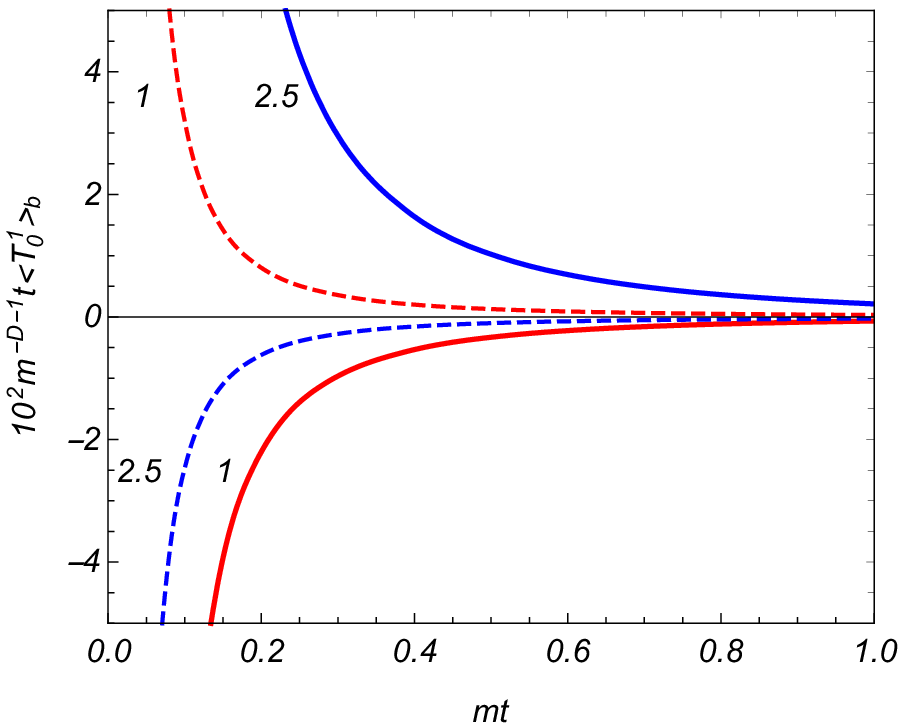,width=7.5cm,height=6.cm}%
\end{tabular}%
\end{center}
\caption{The same as in figure \protect\ref{fig7} for a conformally coupled
field.}
\label{fig8}
\end{figure}

\begin{figure}[tbph]
\begin{center}
\begin{tabular}{cc}
\epsfig{figure=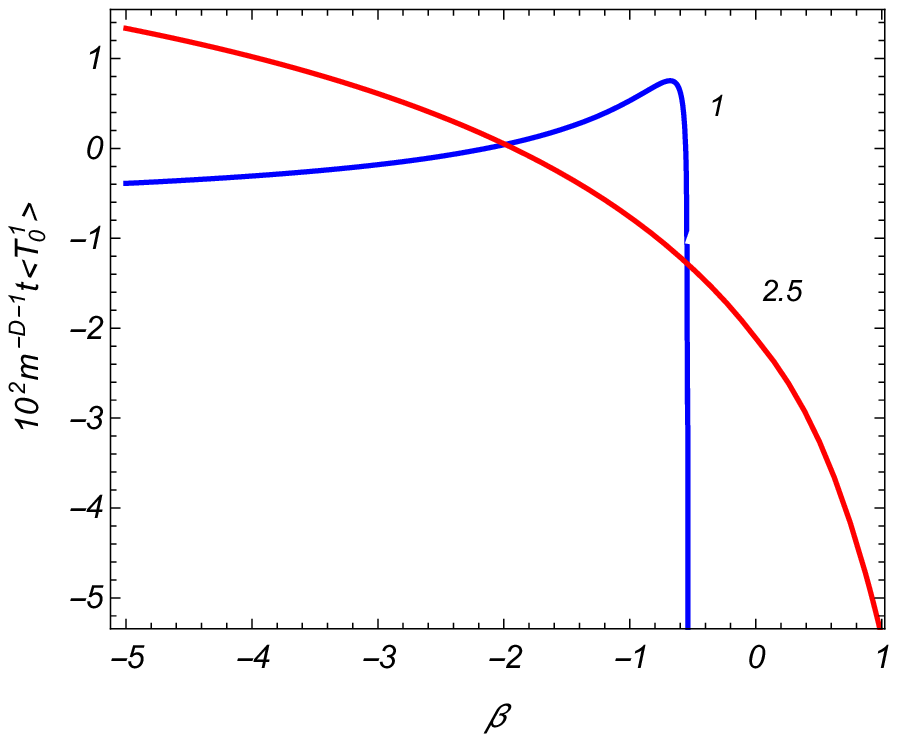,width=7.5cm,height=6.cm} & \quad %
\epsfig{figure=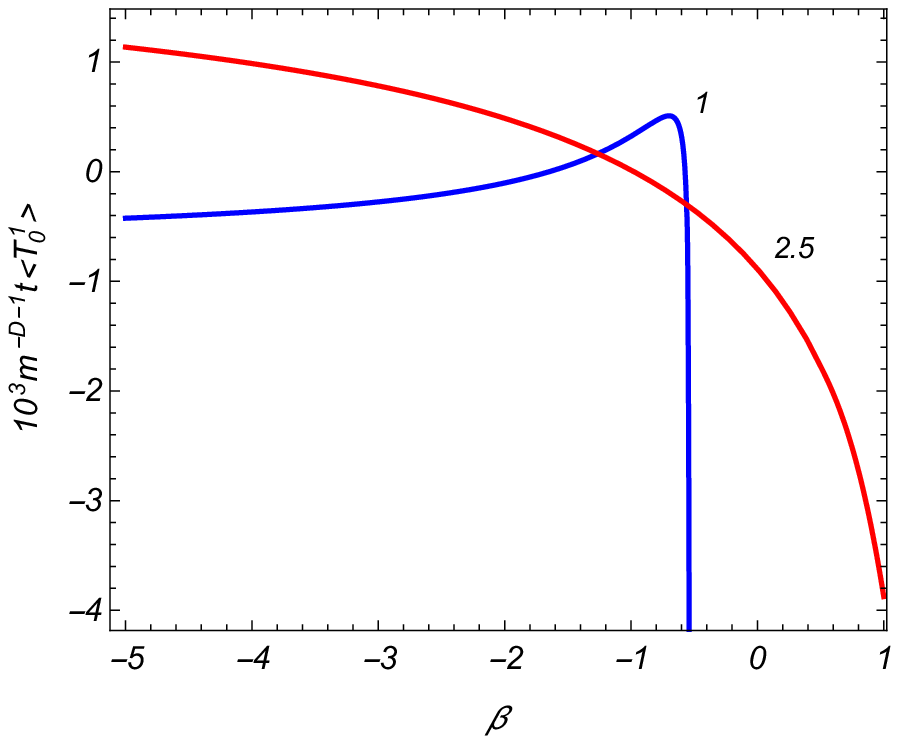,width=7.5cm,height=6.cm}%
\end{tabular}%
\end{center}
\caption{The same as in figure \protect\ref{fig6} for the energy flux
density.}
\label{fig9}
\end{figure}

Having discussed the behavior of the sphere-induced contribution to the VEV
of the energy density, we turn to the investigation of the VEV for the
energy flux. As it has been mentioned above, the energy flux density per
unit proper surface area is given by $t\langle T_{0}^{1}\rangle $. Figure %
\ref{fig7} presents the energy flux density for a spherical boundary in the $%
D=3$ Milne universe and for a minimally coupled scalar field versus the
radial (left panel) and time (right panel) coordinates. The values of the
parameters are the same as those for Figure \ref{fig4}. As it has been
already shown before by the asymptotic analysis, the~energy flux linearly
vanishes at the sphere center. For the values of the parameters
corresponding to the left panel of Figure \ref{fig7} the energy flux is
directed from the sphere for Dirichlet boundary condition and towards the
sphere for Robin boundary condition. For $mt\ll 1$ the leading term in the
asymptotic expansion is given by (\ref{T01sm}) and the energy flux behaves
like $t\langle T_{0}^{1}\rangle \propto 1/t^{4}$ (the right panel in Figure~%
\ref{fig7}). At late stages of the expansion, $mt\gg 1$, the energy flux
exhibits oscillatory damping behavior, $t\langle T_{0}^{1}\rangle \propto
\cos (2mt)/t^{4}$. The same graphs for a conformally coupled field are
displayed in Figure \ref{fig8}. The~qualitative behavior of the energy flux
is similar to that for the case of a minimally coupled field. Now, the
divergence on the sphere is weaker and the corresponding VEV is smaller by
the modulus.
The dependence of the energy flux density on the coefficient in the Robin
boundary condition is depicted in Figure \ref{fig9} for the same values of
the parameters as in Figure \ref{fig6}. As seen from the graphs, in both the
cases of minimally and conformally coupled fields, depending on the value of
the Robin coefficient, the energy flux can be either positive or negative.

\section{Exterior Region}

\label{sec:Ext}

\subsection{Scalar Modes and the Hadamard Function}

In this section we consider the VEVs in the region outside the sphere, $%
r>r_{0}$. As a general solution of the radial equation we take the linear
combination of the associated Legendre functions $P_{iz-1/2}^{-\mu }(u)$ and
$Q_{iz-1/2}^{-\mu }(u)$. One of the coefficients is determined by the
boundary condition (\ref{Robin}) and the second one is found from the
normalization condition. The mode functions realizing the conformal vacuum
are presented as%
\begin{equation}
\varphi _{\sigma }\left( x\right) =C_{\sigma }^{(\mathrm{e})}\frac{X_{iz}(mt)%
}{t^{(D-1)/2}}\frac{Y_{iz-1/2}^{-\mu }(u)}{\sinh ^{D/2-1}r}Y(m_{p};\vartheta
,\phi ),  \label{phie2}
\end{equation}%
where the function $X_{iz}(mt)$ is given by (\ref{Xconf}), $C_{\sigma }^{(%
\mathrm{e})}$ is a normalization constant and%
\begin{equation}
Y_{iz-1/2}^{-\mu }(u)=\bar{Q}_{iz-1/2}^{-\mu }(u_{0})P_{iz-1/2}^{-\mu }(u)-%
\bar{P}_{iz-1/2}^{-\mu }(u_{0})Q_{iz-1/2}^{-\mu }(u).  \label{Ze}
\end{equation}%
The normalization coefficient is determined from the condition (\ref{norm2})
with $Z_{iz-1/2}^{-\mu }(u)=C_{\sigma }^{(\mathrm{e})}Y_{iz-1/2}^{-\mu }(u)$%
. For the region outside the sphere it takes the form%
\begin{equation}
|C_{\sigma }^{(\mathrm{e})}|^{2}\int_{u_{0}}^{\infty }du\,Y_{iz-1/2}^{-\mu
}(u)[Y_{iz^{\prime }-1/2}^{-\mu }(u)]^{\ast }=\frac{\delta (z-z^{\prime })}{%
N(m_{p})}.  \label{nc2}
\end{equation}%
For the evaluation of the integral we note that it diverges for $z^{\prime
}=z$ and the dominant contribution comes from large values of $u$. By using
the corresponding asymptotic expressions for the functions $P_{iz-1/2}^{-\mu
}(u)$ and $Q_{iz-1/2}^{-\mu }(u)$ we can show that
\begin{equation}
\int_{u_{0}}^{\infty }du\,Y_{iz-1/2}^{-\mu }(u)[Y_{iz^{\prime }-1/2}^{-\mu
}(u)]^{\ast }=\frac{\bar{Q}_{iz-1/2}^{-\mu }(u_{0})[\bar{Q}_{iz-1/2}^{-\mu
}(u_{0})]^{\ast }}{|\Gamma \left( 1/2+iz+\mu \right) |^{2}}\frac{\pi \delta
(z-z^{\prime })}{z\sinh (\pi z)}.  \label{IntY}
\end{equation}%
On the base of this, from (\ref{nc2}) for the normalization coefficient one
finds%
\begin{equation}
|C_{\sigma }^{(\mathrm{e})}|^{2}=\frac{z\sinh (\pi z)|\Gamma \left(
1/2+iz+\mu \right) |^{2}}{\pi N(m_{p})\bar{Q}_{iz-1/2}^{-\mu }(u_{0})[\bar{Q}%
_{iz-1/2}^{-\mu }(u_{0})]^{\ast }}.  \label{Ce}
\end{equation}

Note that for a spherical boundary in the static spacetime with a constant
negative curvature space (the corresponding line element is given by (\ref%
{ds2st})) in addition to the modes with real $z$ one may have the modes with
purely imaginary $z$, $z=i\eta $. For those modes the radial part of the
mode functions is given by $\sinh ^{1-D/2}(r)Q_{\eta -1/2}^{-\mu }(u)$. The
allowed values for $\eta $ are determined by the boundary condition on the
sphere and they are roots of the equation $\bar{Q}_{\eta -1/2}^{-\mu
}(u_{0})=0$. As it has been discussed in \cite{Bell14}, for~a given $l$ the
latter equation has no solution for $\beta <\beta _{l}^{(\mathrm{e})}(u_{0})$
with some critical value $\beta _{l}^{(\mathrm{e})}(u_{0})$. The~latter
obeys the conditions $\beta _{l}^{(\mathrm{e})}(u_{0})<\beta _{l+1}^{(%
\mathrm{e})}(u_{0})$, $\beta _{l}^{(\mathrm{e})}(u_{0})>\beta _{l}^{(\mathrm{%
i})}(u_{0})$, and $\beta _{l}^{(\mathrm{e})}(u_{0})>(D-1)/2$. As it has been
discussed above, in the problem at hand with the line element (\ref{ds2})
and for the conformal vacuum the quantum number $z$ should be real. In what
follows we will assume the values of the Robin coefficient in the region $%
\beta <\beta _{0}^{(\mathrm{e})}(u_{0})$, where the equation $\bar{Q}_{\eta
-1/2}^{-\mu }(u_{0})=0$ has no solution.

Substituting the mode functions into the mode sum formula (\ref{WFsum}) with
$\sum_{k=1}^{\infty }$ replaced by the integral $\int_{0}^{\infty }dz$, for
the Hadamard function in the region $r>r_{0}$ we get the representation%
\begin{eqnarray}
G(x,x^{\prime }) &=&\frac{(tt^{\prime })^{(1-D)/2}}{2nS_{D}}%
\sum_{l=0}^{\infty }\frac{\left( 2l+n\right) C_{l}^{n/2}(\cos \theta )}{%
\left( \sinh r\sinh r^{\prime }\right) ^{D/2-1}}\int_{0}^{\infty }dz\,\frac{%
z|\Gamma \left( 1/2+iz+\mu \right) |^{2}}{\bar{Q}_{iz-1/2}^{-\mu }(u_{0})[%
\bar{Q}_{iz-1/2}^{-\mu }(u_{0})]^{\ast }}  \notag \\
&&\times \{J_{-iz}(mt)J_{iz}(mt^{\prime })Y_{iz-1/2}^{-\mu
}(u)[Y_{iz-1/2}^{-\mu }(u^{\prime })]^{\ast }+(\mathrm{c.c.})\},  \label{Ge}
\end{eqnarray}%
where $(\mathrm{c.c.})$ stands for the complex conjugate of the first term
in the figure braces. In order to extract the part in (\ref{Ge}) induced by
the spherical boundary, we subtract the Hadamard function for the
boundary-free geometry, given by (\ref{W0}).

The sphere-induced contribution to the Hadamard function, $G_{\mathrm{b}%
}(x,x^{\prime })$, can be further transformed by using the relation
\begin{eqnarray}
&&\frac{Y_{iz-1/2}^{-\mu }(u)[Y_{iz-1/2}^{-\mu }(u^{\prime })]^{\ast }}{\bar{%
Q}_{iz-1/2}^{-\mu }(u_{0})[\bar{Q}_{iz-1/2}^{-\mu }(u_{0})]^{\ast }}%
-P_{iz-1/2}^{-\mu }(u)P_{iz-1/2}^{-\mu }(u^{\prime })=\frac{-ie^{i\mu \pi }}{%
\pi \sinh (\pi z)}  \notag \\
&&\qquad \times \sum_{s=\pm 1}s\cos [\pi (siz-\mu )]\frac{\bar{P}%
_{iz-1/2}^{-\mu }(u_{0})}{\bar{Q}_{siz-1/2}^{-\mu }(u_{0})}Q_{siz-1/2}^{-\mu
}(u)Q_{siz-1/2}^{-\mu }(u^{\prime }).  \label{RelId}
\end{eqnarray}%
As the next step, in the integral involving the right-hand side of (\ref%
{RelId}) we rotate the contour of integration by the angle $\pi /2$ for the
term with $s=-1$ and by the angle $-\pi /2$ for the term with $s=1$. The
poles $\pm ik$, $k=1,2,\ldots $, are avoided by small semicircles in the
right half-plane with radii $\rho $. In the limit $\rho \rightarrow 0$ the
contributions of the integrals over the semicircles in the upper and lower
half-planes cancel each other and the integral is reduced to the principal
value of the integral over the positive imaginary semiaxis. The product of
the associated Legendre functions is transformed to $Q_{x-1/2}^{-\mu
}(u)Q_{x-1/2}^{-\mu }(u^{\prime })$. By using the relation
\begin{equation}
Q_{x-1/2}^{-\mu }(u)=e^{-2i\mu \pi }\frac{\Gamma (x-\mu +1/2)}{\Gamma (x+\mu
+1/2)}Q_{x-1/2}^{\mu }(u),  \label{Qrel}
\end{equation}%
for the boundary-induced contribution in the Hadamard function we find the
final representation
\begin{eqnarray}
G_{\mathrm{b}}(x,x^{\prime }) &=&-\frac{\left( tt^{\prime }\right) ^{(1-D)/2}%
}{nS_{D}}\sum_{l=0}^{\infty }\frac{\left( 2l+n\right) C_{l}^{n/2}(\cos
\theta )}{(\sinh r\sinh r^{\prime })^{D/2-1}}e^{-i\mu \pi }  \notag \\
&&\times \int_{0}^{\infty }dx\,x\frac{\bar{P}_{x-1/2}^{-\mu }(u_{0})}{\bar{Q}%
_{x-1/2}^{\mu }(u_{0})}V(t,t^{\prime },x)Q_{x-1/2}^{\mu }(u)Q_{x-1/2}^{\mu
}(u^{\prime }).  \label{Wfe4}
\end{eqnarray}%
Note that we consider the values of the Robin coefficient $\beta $ for which
the function $\bar{Q}_{x-1/2}^{\mu }(u_{0})$ in the integrand has no zeros.
Comparing (\ref{Wfe4}) with (\ref{Wb}), we see that the expressions for the
boundary-induced Hadamard functions inside and outside the sphere differ by
the interchange
\begin{equation}
P_{x-1/2}^{-\mu }\rightleftarrows Q_{x-1/2}^{\mu }  \label{ReplPQ}
\end{equation}%
of the associated Legendre functions. Similar to (\ref{Wb}), the integral in
(\ref{Wfe4}) is understood in the sense of the principal value.

The Hadamard functions $G(x,x^{\prime })$ in the interior and exterior regions, given above,
characterize the correlations of vacuum fluctuations at different spacetime points.
Another important global characteristic of the quantum fluctuations correlations is the entanglement entropy.
For a scalar field in two regions of de Sitter spacetime separated by a spherical boundary, the entanglement
entropy has been investigated in \cite{Mald13}-\cite{Chou20}. Note that in those references the de Sitter
hyperbolic open charts were used with the spatial sections similar to (\ref{ds2}). The corresponding mode
functions for scalar fields with general curvature coupling parameter have been discussed in \cite{Sasa95}.
Similar investigations for the entanglement entropy can be done in the Milne Universe for the conformal vacuum
by using the mode functions presented above.

\subsection{VEVs of the Field Squared and Energy-Momentum Tensor}

Taking the coincidence limit $x^{\prime }\rightarrow x$ in (\ref{Wfe4}) we
find the sphere-induced part in the VEV\ of the field squared outside the
sphere:
\begin{equation}
\left\langle \varphi ^{2}\right\rangle _{\mathrm{b}}=-\sum_{l=0}^{\infty }%
\frac{e^{-i\mu \pi }D_{l}}{S_{D}t^{D-1}}\int_{0}^{\infty }dx\,x\frac{\bar{P}%
_{x-1/2}^{-\mu }(u_{0})}{\bar{Q}_{x-1/2}^{\mu }(u_{0})}\frac{F_{\mu }^{(%
\mathrm{e})}(t,r,x)}{\sin (\pi x)},  \label{phi2e}
\end{equation}%
where%
\begin{equation}
F_{\mu }^{(\mathrm{e})}(t,r,x)=J_{x}(mt)J_{-x}(mt)\frac{[Q_{x-1/2}^{\mu
}(u)]^{2}}{\sinh ^{D-2}r}.  \label{Fmue}
\end{equation}%
For a massless field the result (\ref{phi2e}) is conformally related to the
corresponding expression for a conformally coupled massless scalar field in
static spacetime with a negative constant curvature space (see (\ref{phi2m0}%
)). In the exterior region the VEV $\left\langle \varphi ^{2}\right\rangle _{%
\mathrm{b}}^{(\mathrm{st})}$ has the form (\ref{phi2st}) with the
replacements (\ref{ReplPQ}). In the early stages of the expansion, $mt\ll 1$%
, the leading term is given by (\ref{phi2t0}) and the boundary-induced VEV\
behaves as $1/t^{D-1}$. At late stages of the expansion the asymptotic
behavior of $\left\langle \varphi ^{2}\right\rangle _{\mathrm{b}}$ is
described by (\ref{phi2larget}) with the replacements (\ref{ReplPQ}).

Near the sphere the leading term in the asymptotic expansion of the
sphere-induced VEV in the field squared is given by (\ref{phi2near}) with
the replacement $r_{0}-r\rightarrow r-r_{0}$. At large distances from the
sphere, assuming that $r\gg 1$ for fixed $r_{0}$, we can use the asymptotic
expression
\begin{equation}
\frac{Q_{x-1/2}^{\mu }(u)}{\sinh ^{D/2-1}r}\approx 2^{D/2-1}\sqrt{\pi }\frac{%
e^{i\mu \pi }\Gamma (x+1/2+\mu )}{\Gamma (x+1)e^{[x+(D-1)/2]r}}.
\label{Qlarge}
\end{equation}%
The dominant contribution to the integral in (\ref{phi2e}) comes from the
region near the lower limit and to the leading order we get%
\begin{equation}
\left\langle \varphi ^{2}\right\rangle _{\mathrm{b}}\approx -\frac{%
2^{D-3}J_{0}^{2}(mt)}{S_{D}t^{D-1}re^{(D-1)r}}\sum_{l=0}^{\infty }e^{i\mu
\pi }D_{l}\frac{\bar{P}_{-1/2}^{-\mu }(u_{0})}{\bar{Q}_{-1/2}^{\mu }(u_{0})}%
\Gamma ^{2}(l+D/2-1/2).  \label{phi2eLargr}
\end{equation}%
The VEV is exponentially suppressed for both massive and massless fields.
Note that for a spherical boundary in the static geometry described by (\ref%
{ds2st}), at large distances from the sphere the boundary-induced
contribution behaves like (see \cite{Bell14}) $e^{-(2z_{m}+D-1)r}/\sqrt{r}$,
with $z_{m}=\sqrt{m^{2}a^{2}-D(D-1)(\xi -\xi _{D})}$. It is assumed that the
expression under the square root is nonnegative (for negative values the
vacuum state is unstable). For $z_{m}>0$ the decay of the sphere-induced VEV
in the static spacetime is stronger.

\begin{figure}[tbph]
\begin{center}
\epsfig{figure=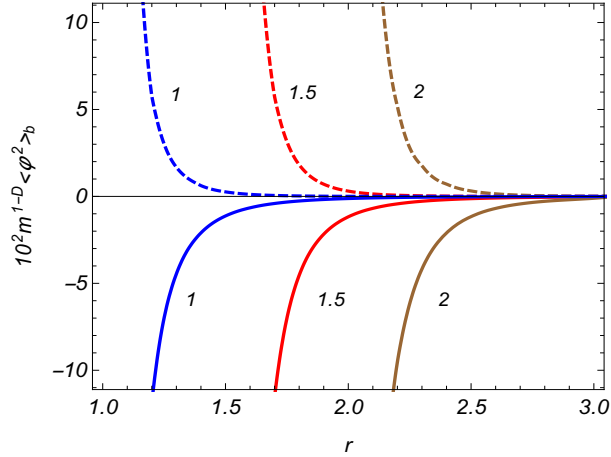,width=8cm,height=6cm}
\end{center}
\caption{The boundary-induced contribution in the VEV of the field squared
outside a sphere for $D=3$ scalar field as a function of the radial
coordinate. The graphs are plotted for $mt=1$, $r_{0}=1,1.5,2$ (numbers near
the curves).}
\label{fig10}
\end{figure}

In Figure \ref{fig10} we have displayed the sphere induced contribution in
the VEV of the field squared outside the spherical shell as a function of
the radial coordinate. The graphs are plotted for $D=3$, $mt=1$, and for the
sphere radius $r_{0}=1,1.5,2$ (the numbers near the curves). The full curves
correspond to Dirichlet boundary condition and the dashed curves present the
case of Robin boundary condition with $\beta =-0.6$. Examples of the
dependence of $\left\langle \varphi ^{2}\right\rangle _{\mathrm{b}}$ on time
and on the Robin coefficient are presented in Figures \ref{fig2} (right
panel) and \ref{fig3}. Note that for the Robin coefficient in the exterior
region we have taken the values $\beta \leqslant (D-1)/2$ for which the
function $\bar{Q}_{x-1/2}^{\mu }(u_{0})$ in the integrand of (\ref{phi2e})
has no zeros.

The VEV of the energy-momentum tensor is found on the base of (\ref{Wfe4})
and (\ref{phi2e}) by using the formula (\ref{EMT}). The calculations are
similar to that for the interior region and for the diagonal components we
get (no summation over $k$)%
\begin{equation}
\langle T_{k}^{k}\rangle _{\mathrm{b}}=\sum_{l=0}^{\infty }\frac{e^{-i\mu
\pi }D_{l}}{S_{D}t^{D+1}}\int_{0}^{\infty }dx\,\frac{x}{\sin (x\pi )}\frac{%
\bar{P}_{x-1/2}^{-\mu }(u_{0})}{\bar{Q}_{x-1/2}^{\mu }(u_{0})}\hat{F}%
_{(k)}F_{\mu }^{(\mathrm{e})}(t,r,x),  \label{Tkke}
\end{equation}%
with the operators $\hat{F}_{(k)}$ from (\ref{F(0)}). The expression for the
VEV of the off-diagonal component reads

\begin{eqnarray}
\langle T_{0}^{1}\rangle &=&\sum_{l=0}^{\infty }\frac{e^{-i\mu \pi }D_{l}}{%
4S_{D}t^{D+2}}\int_{0}^{\infty }dx\,\frac{x}{\sin (\pi x)}\frac{\bar{P}%
_{x-1/2}^{-\mu }(u_{0})}{\bar{Q}_{x-1/2}^{\mu }(u_{0})}  \notag \\
&&\times \left[ 4D\left( \xi -\xi _{D}\right) +\left( 1-4\xi \right)
t\partial _{t}\right] \partial _{r}F_{\mu }^{(\mathrm{e})}(t,r,x).
\label{T01e}
\end{eqnarray}%

The VEVs are connected by the trace relation (\ref{tracerel}) and by the
covariant continuity equations (\ref{Contequ}). For a massless field one has
\begin{equation}
F_{\mu }^{(\mathrm{e})}(t,r,x)=\frac{\sin (\pi x)}{\pi x}\frac{%
[Q_{x-1/2}^{\mu }(u)]^{2}}{\sinh ^{D-2}r},  \label{Fmuem0}
\end{equation}%
and the expressions for the operators $\hat{F}_{(k)}$ are simplified
omitting the terms with time derivatives. In~this special case for the
energy flux one gets%
\begin{equation}
\langle T_{0}^{1}\rangle =\frac{D\left( \xi -\xi _{D}\right) }{\pi
S_{D}t^{D+2}}\sum_{l=0}^{\infty }e^{-i\mu \pi }D_{l}\int_{0}^{\infty }dx\,%
\frac{\bar{P}_{x-1/2}^{-\mu }(u_{0})}{\bar{Q}_{x-1/2}^{\mu }(u_{0})}\partial
_{r}\frac{[Q_{x-1/2}^{\mu }(u)]^{2}}{\sinh ^{D-2}r}.  \label{T01em0}
\end{equation}%
It vanishes for a conformally coupled field. In the latter case the VEVs of
the diagonal components are obtained from those for a spherical boundary in
a static spacetime with negative constant curvature space by conformal
transformation, $\langle T_{k}^{k}\rangle _{\mathrm{b}}=(a/t)^{D+1}\langle
T_{k}^{k}\rangle _{\mathrm{b}}^{(\mathrm{st})}$, where $\langle
T_{k}^{k}\rangle _{\mathrm{b}}^{(\mathrm{st})}$ is given by (\ref{Tkkst})
with the replacements (\ref{ReplPQ}).

Now let us consider the asymptotics at early and late stages of the
expansion. For $mt\ll 1$, in the leading order the function $F_{\mu }^{(%
\mathrm{e})}(t,r,x)$ does not depend on time and for the diagonal components
we find $\langle T_{k}^{k}\rangle _{\mathrm{b}}\approx \left( a/t\right)
^{D+1}\langle T_{k}^{k}\rangle _{\mathrm{b}}^{(\mathrm{st})}$, where $%
\langle T_{k}^{k}\rangle _{\mathrm{b}}^{(\mathrm{st})}$ is the VEV in static
spacetime for a massless field with curvature coupling parameter $\xi $. In
the case of a non-conformally coupled field the leading term in the energy
flux is given by the right-hand side of (\ref{T01em0}). In order to find the
asymptotic of the energy flux for a conformally coupled field one needs the
next to the leading term in the expansion of the function $F_{\mu }^{(%
\mathrm{e})}(t,r,x)$. In this case the leading term in the expansion of the
energy flux at early stages is given by the right-hand side of (\ref{T01sm2}%
) with the replacements (\ref{ReplPQ}). At late stages of the expansion, $%
mt\gg 1$, the~corresponding asymptotics are given by (\ref{Tkklarge})--(\ref%
{T00large}), again, with replacements (\ref{ReplPQ}).

Similar to the interior region, the VEVs (\ref{Tkke}) diverge on the sphere.
The leading terms in the expansions of the energy density and of the
tangential stresses over the distance from the sphere are given by (\ref%
{Tkknear}) with the replacement $r_{0}-r\rightarrow r-r_{0}$. Hence, near
the sphere these components have the same sign for the interior and exterior
regions. The relations (\ref{T01near}) and (\ref{T11near}) between the
energy flux, normal stress and energy density near the sphere remain the
same in the exterior region. Consequently, near the sphere the components $%
\langle T_{0}^{1}\rangle _{\mathrm{b}}$ and $\langle T_{1}^{1}\rangle _{%
\mathrm{b}}$ have opposite signs in the exterior and interior regions. At
large distances from the sphere we use the asymptotic expression (\ref%
{Qlarge}). The main contribution to the sphere-induced VEVs comes from the
region of the integration near the lower limit. For the diagonal components
to the leading order we find (no summation over $k$)%
\begin{equation}
\langle T_{k}^{k}\rangle _{\mathrm{b}}\approx \frac{2^{D-3}\hat{G}%
_{(k)}J_{0}^{2}(mt)}{S_{D}t^{D+1}re^{(D-1)r}}\sum_{l=0}^{\infty }\Gamma
^{2}(l+(D-1)/2)e^{i\mu \pi }D_{l}\frac{\bar{P}_{-1/2}^{-\mu }(u_{0})}{\bar{Q}%
_{-1/2}^{\mu }(u_{0})},  \label{Tkkelarge}
\end{equation}%
where%
\begin{eqnarray}
\hat{G}_{(0)} &=&-\frac{1}{4}t^{2}\partial _{t}^{2}-\left( D\xi -\frac{D-2}{4%
}\right) t\partial _{t}-m^{2}t^{2}+D\left( D-1\right) \left( \xi -\xi
_{D}\right) ,  \notag \\
\hat{G}_{(2)} &=&\left( \frac{1}{4}-\xi \right) t^{2}\partial _{t}^{2}+\left[
\left( D-1\right) \xi -\frac{D-2}{4}\right] t\partial _{t}.  \label{G2}
\end{eqnarray}%
and $\hat{G}_{(1)}=\hat{G}_{(2)}-D\left( D-1\right) \left( \xi -\xi
_{D}\right) $. In a similar way, for the energy flux one gets%
\begin{eqnarray}
\langle T_{0}^{1}\rangle &\approx &-\frac{2^{D-4}\left( D-1\right) }{%
S_{D}t^{D+2}re^{(D-1)r}}\left[ 2D\left( \xi -\xi _{D}\right)
J_{0}(mt)-\left( 1-4\xi \right) mtJ_{1}(mt)\right]  \notag \\
&&\times J_{0}(mt)\sum_{l=0}^{\infty }\Gamma ^{2}(l+(D-1)/2)e^{i\mu \pi
}D_{l}\frac{\bar{P}_{-1/2}^{-\mu }(u_{0})}{\bar{Q}_{-1/2}^{\mu }(u_{0})}.
\label{T01elarge}
\end{eqnarray}%
Note that for a sphere in static spacetime with constant negative curvature
space and for $z_{m}>0$, at large distances the boundary-induced VEVs in the
diagonal components decay as $e^{-(2z_{m}+D-1)r}/\sqrt{r}$. For $z_{m}=0$
the sphere-induced VEVs in the static problem behave like (no summation over
$k$) $\langle T_{k}^{k}\rangle _{\mathrm{b}}\propto e^{-(D-1)r}/r^{2-\delta
_{1}^{k}}$.

\begin{figure}[tbph]
\begin{center}
\begin{tabular}{cc}
\epsfig{figure=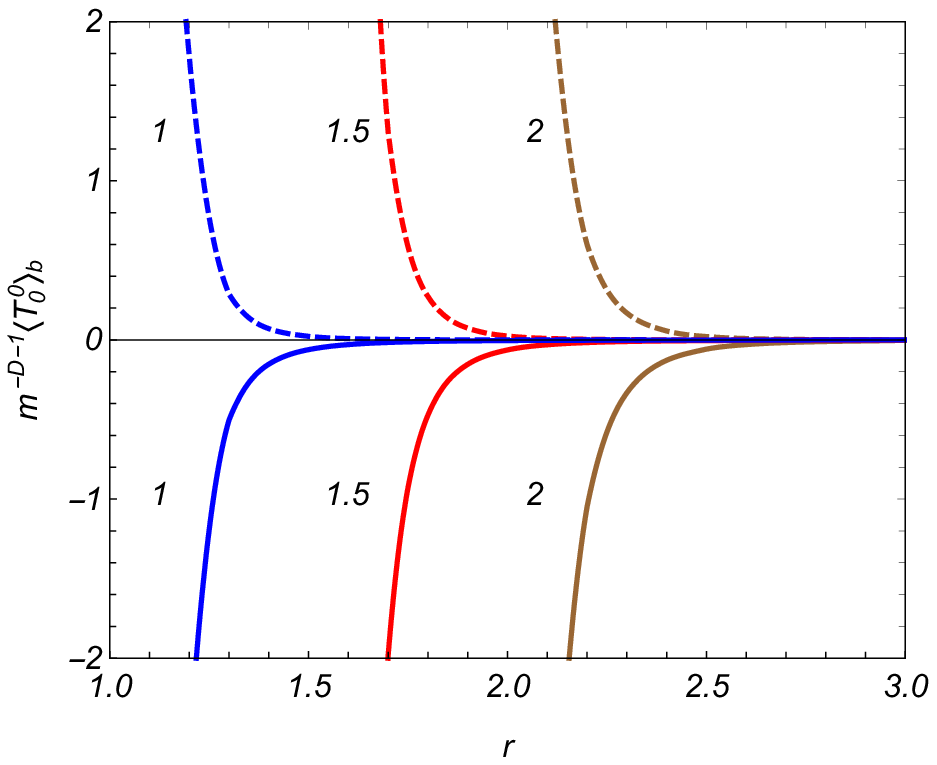,width=7.5cm,height=6.cm} & \quad %
\epsfig{figure=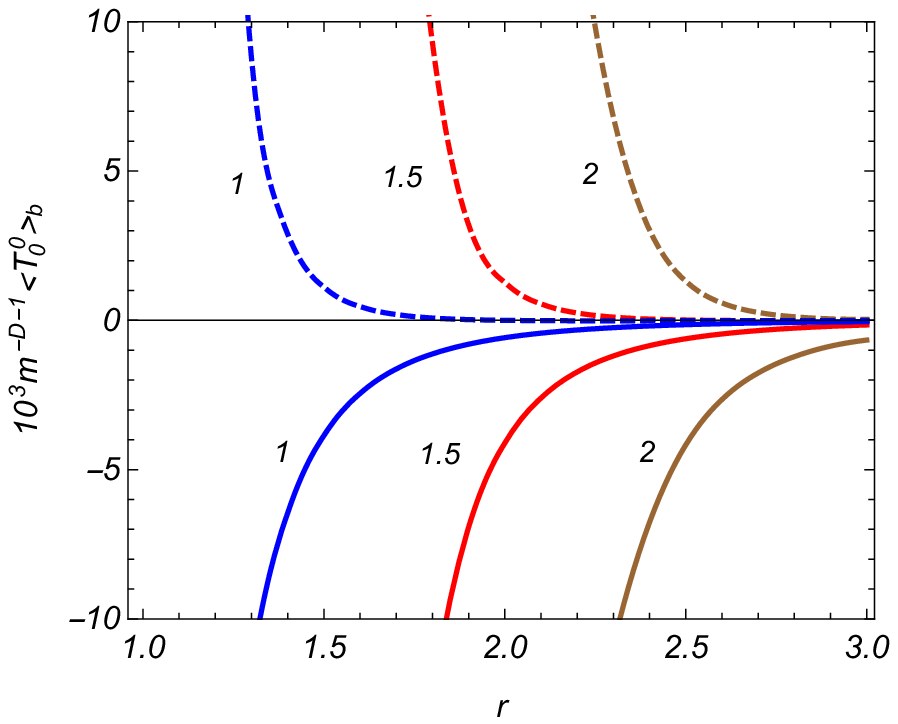,width=7.5cm,height=6.cm}%
\end{tabular}%
\end{center}
\caption{The boundary-induced contributions in the VEV of the energy density
outside the sphere for minimally (left panel) and conformally (right panel)
coupled fields. The graphs are plotted for $D=3$, $mt=1$, and $r_{0}=1,1.5,2$
(numbers near the curves).}
\label{fig11}
\end{figure}

\begin{figure}[tbph]
\begin{center}
\begin{tabular}{cc}
\epsfig{figure=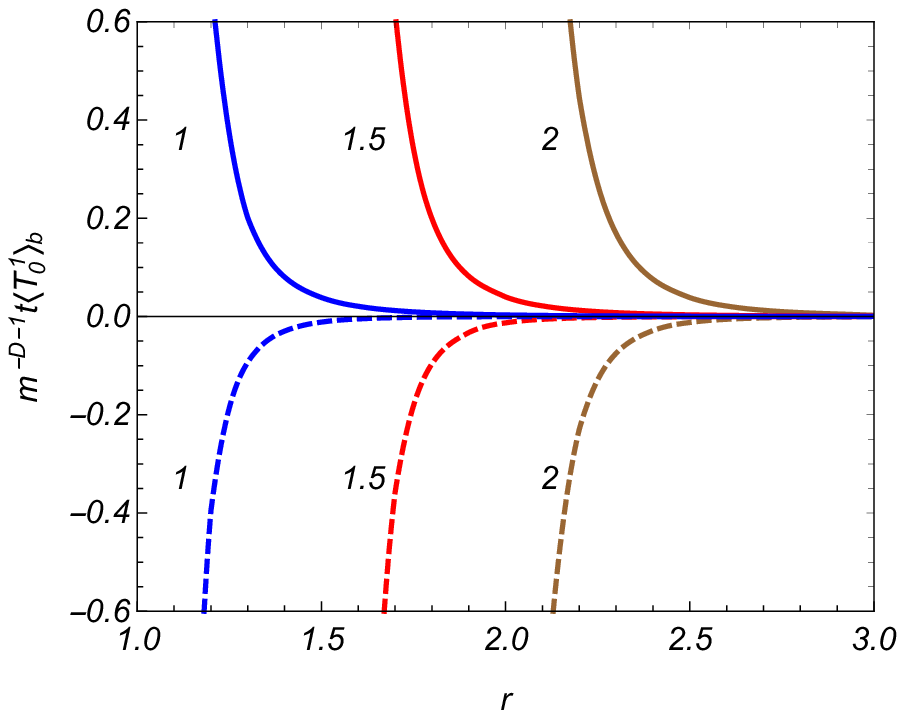,width=7.5cm,height=6.cm} & \quad %
\epsfig{figure=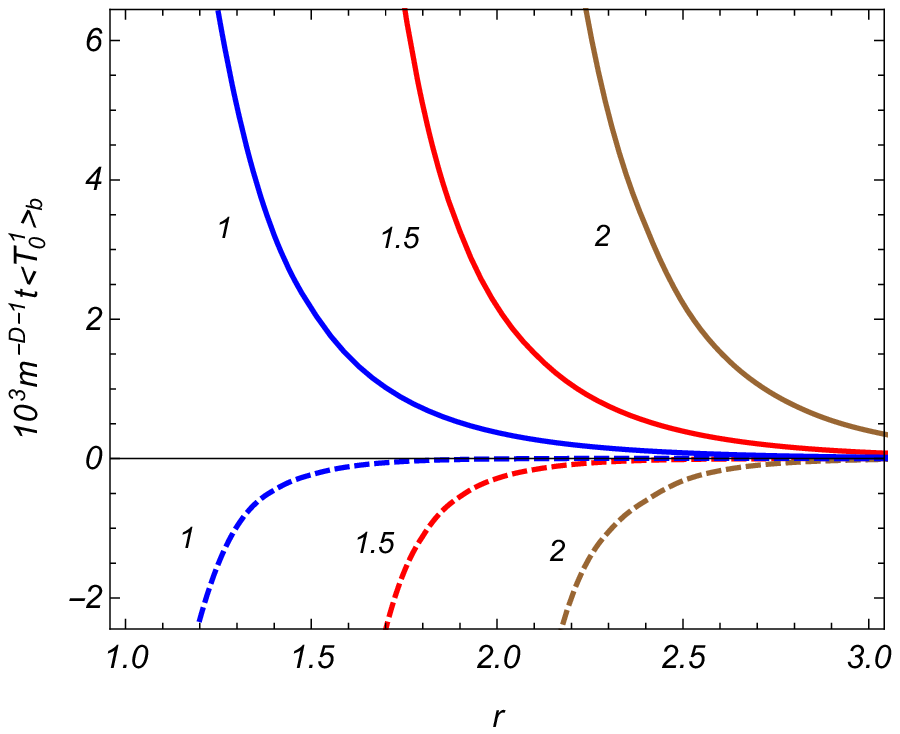,width=7.5cm,height=6.cm}%
\end{tabular}%
\end{center}
\caption{The same as in figure \protect\ref{fig11} for the energy flux
density.}
\label{fig12}
\end{figure}

The figures \ref{fig11} and \ref{fig12} present the radial dependence of the
sphere-induced contributions in the energy density and in the energy flux
density for $D=3$, $mt=1$ and for several values of the sphere radius
(numbers near the curves). The left and right panels correspond to minimally
and conformally coupled fields, respectively. Similar to the interior
region, for the values of the parameters corresponding to figure \ref{fig12}
the energy flux is directed from the sphere for Dirichlet boundary condition
and towards the sphere for Robin boundary condition.

\section{Conclusions}

\label{sec:Conc}

We have investigated the influence of a spherical boundary on the local
properties of the vacuum state for a scalar field in the Milne universe. The
corresponding spacetime metric is flat with spatial sections having a
constant negative curvature and the scale factor is a linear function of
time. In the Minkowskian coordinates, the boundary under consideration
corresponds to a uniformly expanding spherical shell. The mode functions are
found in general case and they are specified for special cases of the
adiabatic and conformal vacua. Further consideration for the two-point
functions and for the VEVs is presented in the case of the conformal vacuum.
Inside the sphere the eigenvalues of the quantum number $z$ are solutions of
the Equation (\ref{Eigeq}) with the barred notation from (\ref{Barnot}).
Depending on the value of the coefficient $\beta $ in Robin boundary
condition, that equation may have purely imaginary roots. By taking into
account that for the conformal vacuum $z$ should be real, we specify the
corresponding region for the values of $\beta $. In the model under
consideration all the information on the properties of the vacuum state is
encoded in two-point functions. As such we have considered the Hadamard
function. The corresponding mode sum for the interior region contains series
over the roots of the eigenvalue Equation (\ref{Eigeq}). For the summation
of the series, a variant of the generalized Abel--Plana formula is used. As a
result of that the contribution of the sphere is explicitly extracted. Among
the advantages of the corresponding integral representation is that the
explicit knowledge of the eigenvalues for the quantum number $z$ is not
required.

Then, on the base of the Hadamard function, we have investigated the VEVs of
the field squared and of the energy-momentum tensor inside the sphere. They
are decomposed into the boundary-free and sphere-induced contributions. The
geometry under consideration is conformally related to the problem with a
spherical boundary in a static spacetime with a constant negative curvature
space and it is explicitly checked that for a conformally coupled massless
scalar field the corresponding VEVs are connected by standard conformal
relations. In that special case the energy flux vanishes. For
non-conformally coupled or for massive fields, a new qualitative feature,
induced by the sphere, is the appearance of the nonzero off-diagonal
component of the vacuum energy-momentum tensor that describes energy flux
along the radial direction. Depending on the values of the parameters, the
flux can be directed either from the sphere or towards the sphere. For a
massless field the time dependence is of the form $1/t^{D-1}$ for the VEV\
of the field squared and of the form $1/t^{D+1}$ for the diagonal components
of the energy-momentum tensor and for the energy flux $t\langle
T_{0}^{1}\rangle $.

The general formulas for the VEVs are rather complicated and, in order to
clarify their behavior, we have considered various asymptotic regions of the
parameters. Near the sphere the boundary-induced contributions dominate in
the total VEVs and the leading terms in the corresponding asymptotic
expansions for the VEV of the field squared, for the energy density and for
tangential stresses are the same as those for a sphere in the Minkowski
spacetime (with the distance from the sphere replaced by the proper distance
in the Milne universe). For the normal stress and energy flux the divergence
on the sphere is weaker and the leading terms are given by (\ref{T01near})
and (\ref{T11near}). At early stages of the expansion, $t\rightarrow 0$, the
leading terms in the corresponding asymptotics for the VEVs of the field
squared and of the diagonal components of the energy-momentum tensor are
conformally related to the VEVs in static spacetime with a constant negative
curvature space. The VEV of the field squared behaves as $1/t^{D-1}$ and the
diagonal components behave like $1/t^{D+1}$. The asymptotic of the energy
flux at early stages is given by (\ref{T01sm}). At late stages of the
expansion and for massive fields, assuming that $mt\gg 1$, the~VEVs exhibit
damping oscillatory behavior. The corresponding asymptotics are given by (%
\ref{phi2larget}) for the field squared and by (\ref{Tkklarge})--(\ref%
{T00large}) for the energy-momentum tensor. All these features are displayed
by numerical examples. In particular, we have demonstrated that, depending
on the Robin coefficient, both the sphere-induced energy density and the
energy flux can be either positive or negative.

In the region outside the sphere the eigenvalues of the quantum number $z$
are continuous and the mode functions are given by (\ref{phie2}). The
integral representations for the sphere-induced contributions to the
Hadamard function and to the VEVs of the field squared and energy-momentum
tensor differ from the corresponding expressions for the interior region by
the replacement (\ref{ReplPQ}) of the associated Legendre functions. For the
corresponding formulas the range of the allowed values for the Robin
coefficient is wider compared to that for the interior region. Near the
sphere and for non-conformally coupled fields the VEVs of the field squared,
of the energy density and of the tangential stresses have the same sign in
the exterior and interior regions. The VEVs of the normal stress and of the
off-diagonal component have opposite signs. At large distances from the
sphere the boundary-induced VEVs decay exponentially, like $e^{-(D-1)r}$ for
both massive and massless fields. For massive conformally and minimally
coupled fields and for a spherical boundary in a static spacetime with
constant negative curvature space the decay at large distances is stronger.

\appendix

\section{Summation Formula}

\label{sec:appSF}

The VEVs of physical observables inside a spherical shell in the Milne
universe are expressed in terms of the series of the type $%
\sum_{k=1}^{\infty }T_{\mu }(z_{k},u)h(z_{k})$, where $z_{k}$ is the $k$th
positive root of the Equation (\ref{Eigeq}). A summation formula for those
series with a function $h(z)$ analytic in the right-half plane $\mathrm{Re}%
\,z>0$, has~been derived in \cite{Bell14} (see also \cite{Saha08} for the
case $B=0$). Additional conditions imposed on the function $h(z)$ were
formulated in \cite{Bell14}. In particular, it was assumed that the function
$h(z)$ has no poles on the imaginary axis. In the physical problem we
consider in the present paper the corresponding function has simple poles $%
z=\pm ik$, $k=1,2,\ldots $, at the zeros of the function $\sinh (z\pi )$ in (%
\ref{W}). The~generalization of the summation formula from \cite{Bell14} for
the case of functions $h(z)$ having simple poles on the imaginary axis is
straightforward.

If the function $h(x)$ is real for real $x$, then one has $\left[ h(ix)%
\right] ^{\ast }=h(-ix)$. From here it follows that if $ix$ is a pole of the
function $h(z)$ then $-ix$ is also a pole. Hence, the poles on the imaginary
axis are of the form $\pm ix_{k}$, $x_{k}>0$. The procedure to obtain the
summation formula from the generalized Abel--Plana formula is similar to that
used in \cite{Bell14}. The difference is that now the poles $\pm ix_{k}$
should be avoided by small semicircles with radii $\rho $ in the right
half-plane and with the centers at the points $\pm ix_{k}$. In~the limit $%
\rho \rightarrow 0$ the contributions of the integrals over those contours
are expressed in terms of the corresponding residues and the integral over
the imaginary axis should be understood in the sense of the principal value.
The summation formula takes the form%
\begin{eqnarray}
&&\sum_{k=1}^{\infty }T_{\mu }(z_{k},u_{0})h(z_{k})=\sum_{k}\cos [\pi (\mu
-x_{k})]\frac{\bar{Q}_{x_{k}-1/2}^{-\mu }(u_{0})}{\bar{P}_{x_{k}-1/2}^{-\mu
}(u_{0})}\sum_{j=\pm 1}\mathrm{Res}_{z=jix_{k}}h(z)  \notag \\
&&\qquad +\frac{e^{-i\mu \pi }}{2}\int_{0}^{\infty }dx\,\sinh (\pi x)h(x)-%
\frac{1}{2\pi }\int_{0}^{\infty }dx\,\frac{\bar{Q}_{x-1/2}^{-\mu }(u_{0})}{%
\bar{P}_{x-1/2}^{-\mu }(u_{0})}\cos [\pi (x-\mu )]\sum_{j=\pm }h(xe^{j\pi
i/2}).  \label{SumForm}
\end{eqnarray}%
The remaining conditions on the function $h(z)$ are the same as in \cite%
{Bell14}. Namely, it is assumed that for $z=x+iy$ it obeys the condition $%
|h(z)|<\varepsilon (x)e^{cy\,\mathrm{arccosh}u_{0}}$ for $|z|\rightarrow
\infty $, where $c<2$ and $\varepsilon (x)e^{\pi x}\rightarrow 0$ for $%
x\rightarrow +\infty $.

\end{document}